\documentclass[fleqn,usenatbib]{mnras}

\usepackage{newtxtext,newtxmath}

\usepackage[T1]{fontenc}

\DeclareRobustCommand{\VAN}[3]{#2}
\let\VANthebibliography\thebibliography
\def\thebibliography{\DeclareRobustCommand{\VAN}[3]{##3}\VANthebibliography}

\usepackage{graphicx}	% Including figure files
\usepackage{amsmath}	% Advanced maths commands
\usepackage{amssymb}	% Extra maths symbols
\newcommand{\new}[1]{{#1}}
\newcommand{\mar}[1]{{#1}}
\newcommand{\rvw}[1]{{#1}}

\title[PNe as tracers of stellar populations]{Planetary nebulae as tracers of stellar population properties: a pilot study with MUSE}

\author[A. I. Ennis et al.]{
Ana In\'es Ennis,$^{1,2}$\thanks{mail: ennis.ana@gmail.com}
Johanna Hartke,$^{3,4,5}$, Fuyan Bian$^{6,7}$, Claudia Pulsoni$^{8}$, Chiara Spiniello$^{9,10}$, \newauthor Magda Arnaboldi$^{11}$, Roberto de Propris$^{3,4,12}$
\\
$^{1}$Perimeter Institute for Theoretical Physics, Waterloo, Ontario N2L 2Y5, Canada\\
$^{2}$Waterloo Centre for Astrophysics, University of Waterloo,
Waterloo, Ontario, N2L 3G1, Canada\\
$^{3}$Finnish Centre for Astronomy with ESO (FINCA), FI-20014 University of Turku, Finland\\
$^{4}$Tuorla Observatory, Department of Physics and Astronomy, FI-20014 University of Turku, Finland\\
$^{5}$Turku Collegium for Science, Technology and Medicine (TCSMT), FI-20014 University of Turku, Finland\\
$^{6}$European Southern Observatory, Alonso de Córdova, Vitacura, Santiago, Chile\\
$^{7}$Chinese Academy of Sciences South America Center for Astronomy, National Astronomical Observatories, CAS, Beijing 100101, China\\
$^{8}$Max-Planck-Institut für Extraterrestrische Physik (MPE), Gießenbachstraße 1,
85748 Garching bei München, Germany\\
$^{9}$Department of Physics, University of Oxford, Denys Wilkinson Building, Keble Road, Oxford OX1 3RH, UK\\
$^{10}$INAF – Osservatorio Astronomico di Capodimonte, Salita Moiariello 16, 80131 - Napoli, Italy\\
$^{11}$European Southern Observatory, Karl-Schwarzschild-Straße 2, 85748 Garching bei M\"unchen, Germany\\
$^{12}$ Department of Physics and Astronomy, Botswana International University of Science and Technology, Private Bag 16, Palapye, Botswana}

\date{Accepted XXX. Received YYY; in original form ZZZ}

\pubyear{2015}

\begin{document}
\label{firstpage}
\pagerange{\pageref{firstpage}--\pageref{lastpage}}
\maketitle

\begin{abstract}
Planetary nebulae (PNe) are the only single stars in galaxies outside the Local Group that can be used as kinematic tracers of the diffuse light in the extended halo. Analysing their luminosity-specific number density across galaxies of different morphologies has also shown hints that they may be used as tracers of the age and metallicity of stellar populations. A proper understanding of this relation has been hindered by the fact that simultaneously detecting PNe and accurately measuring stellar properties is extremely difficult using classical narrow-band imaging methods, which cannot detect PNe in the bright centres of galaxies. In this work, we use integral-field spectroscopy to overcome this challenge, analysing the inner regions of a sample of ten early-type galaxies from the Extended Planetary Nebulae Survey (ePN.S) for which archival MUSE data was available. With the Diffuse Emission-Line Filter (DELF) technique, we automate the detection of PNe, and perform spectral fitting on the diffuse light to infer kinematics and stellar population parameters. We compare the PN number density profile and its associated $\alpha$-parameter with multiple properties of the host galaxies. We find that our sample follows the previously observationally constrained correlation with the metallicity of the host galaxy. We find a weak anti-correlation between the $\alpha$-parameter and the FUV excess, highlighting the possible relation between the visibility lifetime of PNe on the spectral energy distribution of their host galaxies, with fewer PNe detected in association with stellar populations characterized by a UV excess. 
\end{abstract}


%%%%%%%%%%%%%%%%%%%%%%%%%%%%%%%%%%%%%%%%%%%%%%%%%%

%%%%%%%%%%%%%%%%% BODY OF PAPER %%%%%%%%%%%%%%%%%%

\section{Introduction}

Towards the end of their lives, stars with initial masses between $\sim 1$ and $8\,\textrm{M}_{\odot}$ go through the planetary nebula (PN) phase, \mar{instead of exploding as supernovae}. During this stage, the central star is surrounded by an envelope of material where recombination and photoionization are happening simultaneously. This creates the distinct optical spectral features of a PN: a flat continuum with prominent emission lines, with \rvw{one of} the strongest emissions being at the 5007\,\AA\ forbidden [\ion{O}{iii}] line. This line is intense enough to be singularly detected from individual PNe at large distances ($\sim 100\,$Mpc, \citealt{Gerhard2005}). This has led PNe to become one of the most widely used kinematic tracers for the stellar populations in galaxy haloes \citep[e.g.][]{Coccato2009,Pulsoni2018,Hartke2022,Bhattacharya2023,Pulsoni2023}, serving as tools to unveil tidal substructures and to further our understanding of dark matter halos. PNe are also viable kinematic tracers of the diffuse intracluster light \citep[e.g.][]{Arnaboldi1996,Gerhard2005,Longobardi2015,Hartke2018,2022FrASS...972283A}.   

PN populations in nearby early- and late-type galaxies (LTGs) have been shown to follow a near-universal luminosity function \citep{Ciardullo2002, Ciardullo2004}, \new{generally well described within 2.5 magnitudes} by
\begin{equation}
    N(M_{5007})=c_{1}e^{c_{2}M_{5007}}(1-e^{3(M^{*}-M_{5007})}),
    \label{eq:pnlf}
\end{equation}
where $M_{5007}$ is the absolute magnitude of the $5007$\,\AA\ line, $c_{1}$ is a normalisation constant, and $M^{*}$ is the absolute magnitude of the bright cut-off, which has empirically been shown to be nearly-constant at $M^{*}=-4.51^{+0.02}_{0.04}$ \citep{Ciardullo2002,Ciardullo2004}. The shape of the analytic PNLF is motivated by the physics of PNe, with the first term describing how the surface brightness of the envelope around the central star is fading as it expands with constant velocity \citep{Henize1963}. The slope $c_2 = 0.307$, considered constant in the original implementation of the function, has since been found to vary.  \cite{Longobardi2013} was the first to  introduce a PNLF with variable faint-end slope, which is thought to be connected to the properties of the underlying stellar populations. Steeper PNLFs have been associated with older, metal-poor stellar populations, while flatter slopes may be connected to younger stars \citep{Longobardi2013,Ciardullo2004,Ciardullo2010,Hartke2020}. \new{In some star-forming galaxies, the PNLF has been shown to deviate from this equation for faint PNe, presenting dips \citep{Jacoby2002,Hernandez-Martinez2009}}. Due to the universality of the bright cut-off, the PNLF is used to estimate extragalactic distances \citep{Ciardullo1989,Buzzoni2006,Jacoby2024}. Recently, \cite{Valenzuela2025} have started to analyse PNe within cosmological simulations, looking to uncover the physical pathway for the PNLF, its shape and its cut-off.

The luminosity-specific PN number density, commonly known as the $\alpha$-parameter, \mar{estimates} the size of the PN population in a galaxy in relation to its luminosity \citep{Jacoby1989}, 
\begin{equation}
    N_{\rm PN}=\alpha L_{\rm bol}.
\end{equation}
\mar{Depending on the depth/sensitivity of the survey, $\alpha_{\Delta m}$ refers to the limit in fainter magnitudes below the bright cut-off magnitude of the PNLF reached by the survey}. The total number of PNe, $N_\mathrm{PN}$, for a chemically uniform, single-age stellar population can be described in terms of the specific evolutionary flux, $\mathcal{B}$, and the PN visibility lifetime, $\tau_\mathrm{PN}$:
\begin{equation}
    N_{\rm PN}=\mathcal{B}\,L_{\rm bol}\,\tau_{\rm PN}
    \end{equation}
In the simple stellar populations (SSP) framework \citep{Renzini1986, Buzzoni1989}, $\mathcal{B}$ is defined as the rate of stars leaving the main sequence normalized by the unit light. The $\tau_{\rm PN}$ refers to the PN visibility lifetime, which is not the same as the time spent in the PN evolutionary phase. In particular, $\tau_{\rm PN}$ depends heavily in the properties of the ejected material and the core-mass evolution of the central star. Therefore, the $\alpha$-parameter links the \mar{important PN phase parameters} with \mar{some of the average }properties of their parent stellar component.
%the properties of the underlying stellar component to those of the individual stars that make up the PNe population.

Observationally, the $\alpha$-parameter has been found to correlate with the colour and morphology of the galaxy, as described initially in \cite{Buzzoni2006}, B06 from here on. This seminal work thoroughly investigated the $\alpha$-parameter-galaxy connection through the use of population synthesis models, comparing theoretical results with a diverse observational sample. They calculate the $\alpha$-parameter based on the single stellar population (SSP) theory derived by \cite{Renzini1986} and \cite{Buzzoni1989}, taking into consideration the chemical and dynamical properties of PN and stellar evolution tracks, and applying them to template galaxy models. The main theoretical prediction from this analysis is that the $\alpha$-parameter should moderately \new{increase} from  star-forming/blue galaxies to quiescent/red ones, with the former hosting smaller PNe populations and a having lower mass/luminosity ratios. \new{Observationally, the earliest work to investigate the empirical correlations is \cite{Ciardullo1991}, where for a sample of ten galaxies the $\alpha$-parameter is shown to be strongly correlated with the U-V index, showing the opposite trend than the one predicted by theory, with blue galaxies producing more PNe. Studies over the following decade have continued to corroborate this result, contrary to the theoretical models, with improved techniques and more accurate measurements, in particular \cite{Ciardullo2005} where the analysed correlations include optical colour of the host galaxy, the UV upturn, and the H$\beta$ absorption line, used as an indicator of the age of the population. }
%This is not seen observationally, where bluer galaxies show higher than average $\alpha$-parameter values, and redder ones present a large scatter in $\alpha$, although it remains consistently lower than in the star-forming galaxies.

In recent works, the $\alpha$-parameter has been shown to vary across different regions of a single galaxy, which may be linked to changes in the age and metallicity of the stellar populations \citep{Longobardi2013,Hartke2017,Bhattacharya2019,Hartke2020}. This dependence is promising since it could turn PNe into tracers of stellar population properties of stellar haloes, and, in the cases of cluster galaxies, the surrounding intracluster light. %However, these $\alpha$-parameter variations are also contentious since they are in principle at odds with the proposed universality of the PNLF. 
\new{These $\alpha$-parameter variations} have implications for our understanding of stellar evolution, since the $\alpha$-parameter is related to the visibility lifetime ($\tau_{\rm{PN}}$) of PN, which depends on properties of the central stars and their ejected material \citep{Renzini1986,Buzzoni1989,Buzzoni2006,Annibali2017}. 

% This parameter is connected to the metallicity of the underlying population, as seen in Figure\,\ref{fig:buzzoni}, where we see the sample of galaxies analysed by B06 indicated with filled symbols in a blue gradient, where circles, squares, and diamonds represent elliptical, spiral and irregular galaxies respectively. In this plot, elliptical galaxies appear to present an approximately constant value of $\alpha$, while irregular and spirals seem to have a gradient of values. In \cite{Hartke2020}, the analysis of the $\alpha$ parameter for the inner and outer region of M105 results in two significantly different values, which seems to be connected to the inner region presenting a xx metallicity. Therefore, we can infer an empirical relation, but a statistically significant sample of galaxies with a known $\alpha$ parameter both in the outer and inner region is crucial to untangle the origin of this connection. 

The main challenge in understanding the possible relation between stellar age and metallicity and the $\alpha$-parameter lies in the fact that PNe are commonly studied in regions where the properties of stellar populations cannot be easily measured. The central regions of galaxies, where direct measurements of stellar population properties are easily obtained, are thought to have distinct formation pathways from the extended halo, where PNe are traditionally observed. Physical properties thus vary with radius across galaxies, in particular age and metallicity \citep{Coccato2010,Pastorello2014,Zibetti2020,Cheng2024}. This means that simultaneously analysing the stellar component and the $\alpha$-parameter of PNe in the same regions of galaxies for a significant sample of galaxies is a key step in unlocking the physical reasons behind the apparent correlation. 

% Because of the different origin of central regions and stellar ha-los, galaxies are expected to show significant variation of phys-ical properties with radius, such as shapes of the light profiles(Huang et al. 2013; D’Souza et al. 2014; Spavone et al. 2017;Buitrago et al. 2017), stellar populations (Coccato et al. 2010;Pastorello et al. 2014; Zibetti et al. 2020), and kinematics (Coc-cato et al. 2009; Romanowsky & Fall 2012; Arnold et al. 2014;Foster et al. 2016) (from Pulsoni et al)

Classical detection methods such as the on-off technique, which is based on narrow-band imaging covering the [\ion{O}{iii}] emission (``on'') in comparison with broad-band imaging away from the [\ion{O}{iii}] emission (``off''), suffer from blind spots in the central regions of galaxies, where the galaxy outshines the PNe\mar{\footnote{In classical narrow-band imaging, the Possionian noise of the continuum in a few tens \AA\ band pass is of the same order of the flux from the brightest PN}}. In the last decade, integral field spectroscopy (IFS)
has emerged as a tool to untangle emission in these regions, both with instruments on 4-meter telescopes \citep[e.g.][]{Sarzi2011}, and with MUSE on the 8.2-meter VLT \citep[e.g.][]{Kreckel2017,Roth2018}. Based on the classical on-off technique, \cite{Roth2021} developed a method called ``DELF'' (differential emission-line filter), which facilitates the detection of emission-line sources in high surface brightness regions of galaxies onto which we build in this paper.
%The goal of their work is to improve the accuracy of the measurement of the PNLF. 
%The DELF method makes use of the high spectral resolution in MUSE data cubes to extract images with a very narrow spectral range, which reduces the Poissonian noise from the background emission due to the galaxy's continuum. Splitting the narrow-band filter in this manner increases the signal-to-noise ratio (SNR) of the emission lines, facilitating the detection of fainter objects even in the central regions. %Historically, the PNLF was used to measure distances up to 15\,Mpc, whereas DELF allows for this to be extended to distances as large as 40\,Mpc, hence making the PNLF a potential tool to constrain the Hubble constant too \citep{Jacoby2024}.   

In this work we aim to detect PNe in the inner regions of ETGs in order to connect their properties with that of PNe located further from the centre. \mar{We use archival data of ten ETGs located in dense environments to detect and characterize PNe, and, at the same time, obtain the properties of the parent stellar population, such as age and metallicity. 
By exploring the potential of MUSE data in the ESO Science Archive, we correlate the PN samples with the accurate and spatially resolved measurements of the stellar population properties in the bright centres of ETGs which are allowed by the IFS \citep{Kuntschner2010,McDermid2015}. 
The goals are to allow us to correlate the $\alpha$-parameter and other parameters all measured in the same spatial regions. } 

\begin{figure*}
    \centering
    \includegraphics[width=\linewidth]{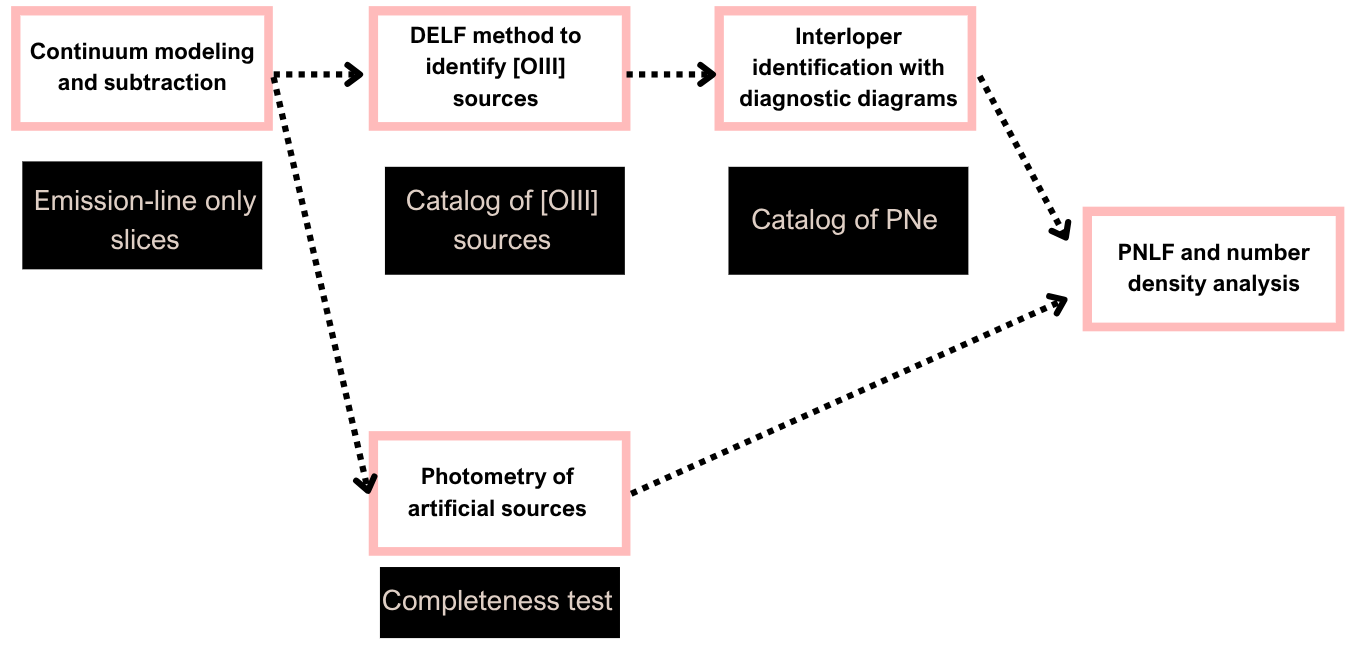}
    \caption{Flowchart describing the method used to perform the analysis carried out in this paper.}
    \label{fig:flowchart}
\end{figure*}

This paper is organised as follows. In Section~\ref{sect:obs} we present the sample of archival MUSE observations. In Section~\ref{sect:stellar_pops} we derive spatially resolved stellar kinematics and population properties. In Sections~\ref{sect:PN_detection} and \ref{sect:PN_analysis} we present our automatised PN detection method and subsequent analysis of the PN samples, shown in Figure\,\ref{fig:flowchart} for reference. Finally, in Section~\ref{sect:discussion}, we discuss our results and contrast PN and stellar population properties. We conclude the paper in Section~\ref{sect:conclusions}. Throughout the paper, we adopt distances based on surface brightness fluctuations (SBF) from \cite{Tully2013} \new{(compilation based on the sample from \cite{Tonry2001} with modifications by \cite{Blakeslee2010})} for the selected galaxies. \new{There is a zero-point systematic of about $0.2$\, mag between PNLF and SBF distances, which will affect our bright cut-off measurements but since it is homogenously applied throughout the sample it does not impact the analysis.} %, and we consider the $\alpha$-parameter as $\alpha_{2.5}$, meaning our magnitudes are between the bright cut-off, $m_{*}$, and $m_{*}+2.5$. This is done for homogeneity, since the completeness levels vary through the sample, and to facilitate future comparisons.

\section{Data reduction and sample selection}
\label{sect:obs}

\subsection{Observations and data reduction}
The observations were obtained from the Science Portal in the ESO Archive, where processed data are readily available \mar{from the MUSE science data collection (https://doi.eso.org/10.18727/archive/41)}. All data cubes used in this work were observed using the MUSE spectrograph \citep{Bacon2010}, located at the Nasmyth B focus of Yepun, the 8.2\,m VLT UT4 telescope. The observations in every case were made in wide-field mode, covering a $1\times1$ field of view (FoV) with a pixel scale of 0.2''. The spectral coverage of MUSE is $\sim4800-9300$\,\AA, with a spectral sampling of $1.25$\,\AA, and a spectral resolution of $\sim 2.6$\,\AA. The data products were obtained from the ESO Archive Science Portal, and they were produced by ESO with the standard pipeline \citep{Weilbacher2020}. The exception were the comissioning data for NGC\,3115, which we reduced ourselves with the standard MUSE data reduction pipeline and the dedicated calibrations for this commissioning run since only raw data were available. 

\subsection{Sample selection}
We built our sample of galaxies, listed in Table\,\ref{tab:sample}, based on the extended Planetary Nebula Spectrograph (ePN.S) early-type galaxy survey \citep{Pulsoni2018} and the Fornax Cluster VLT Spectroscopic Survey \citep{Spiniello2018}, selecting those for which MUSE \mar{science} data was available in the ESO archive\footnote{\url{https://archive.eso.org/scienceportal/home}}, with the goal of analysing the $\alpha$-parameter \mar{properties} across entire galaxies as future work. We discarded those with cubes that lacked the required depth due to limited exposure times. The sample contains galaxies located up to \new{a distance of} $25$\,Mpc, a constraint inherited from the ePN.S sample. Due to the range of distances, the variety of pointings and the presence of dust or emission sources, we cover different galactocentric distances in each galaxy. 

All galaxies in our sample are massive ($10<\rm{log}\,M_{gal}/M_{\odot}<11.5$) ETGs, meaning their stellar populations are old, and mainly \new{metal-rich}, which constrains the range of analyzed properties in comparison with B06 which includes late-type galaxies. In the following we will use the galaxy NGC\,1387 as a test case, describing the analysis which is performed on all the ETGs in the sample. The results for the rest of the sample is shown in Appendix\,\ref{sec:app1} and Appendix\,\ref{sec:app2}.

\begin{table*}
	\centering
	\caption{Galaxies included in the sample. Asterisks indicate galaxies for which observations with different exposure times were considered. All distances are from \protect \cite{Tully2013}, \new{where measurements are compiled from \protect \cite{Tonry2001}, with small modifications from \protect \cite{Blakeslee2010}}. Other observations for NGC\,3115 currently available in the Archive were not used due to the seeing conditions being inadequate for this work. Morphologies are from $^{[1]}$: \protect \cite{Kartha2014} and $^{[2]}$: \protect \cite{deVaucouleurs1991}}
	\label{tab:sample}
	\begin{tabular}{lccccccc}
		\hline
        Galaxy & Distance [Mpc] & Redshift & Morph. type & m$_{B}$ & Total exposure time [s] & \new{Seeing [arcsec]} & ESO programme ID \\
        \hline
        % NGC 1316 & 17.5 & 0.0060 & E & 0 & 1700 & Walcher, C.J. & 094.B-0298\\
        NGC 1387 & 19.1 & 0.0043 & S0$^{[1]}$ & 11.75 & 3600 & 0.82 & 296.B-5054\\
        NGC 1399 & 22.1 & 0.0047 & E1$^{[2]}$ & 9.74 & 7155 & 0.89 & 094.B-028 \& 094.B-0903\\
        NGC 1404 & 19.5 & 0.0065 & E1$^{[2]}$ & 10.69 & 3600 & 0.83/0.89 & 296.B-5054\\
        NGC 3115 & 9.5 & 0.0023 & S0 & 11 & 600 & 0. & 60.A-9100(A)\\
        NGC 3379 & 10.2 & 0.001825 & E1$^{[2]}$ & 10.56 & 1656 & 0. & 108.21ZY\\
        NGC 4365 & 22.8 & 0.0041 & E3$^{[2]}$ & 11.5 & 2502 & 1.03 & 094.B-0225\\
        NGC 4374 & 16.9 & 0.0034 & E1$^{[2]}$ & 12.09 & 1200 & 0.68 & 0102.B-0048\\
        NGC 4472 & 16.1 & 0.0033 & E2$^{[2]}$ & 13.21 & 1800 & 0.88 &095.B-0295\\
        NGC 4594 & 11.3 & 0.0034 & E2$^{[2]}$ & 9.55 & 2700-3600* & 1.03 - 1.43 & 60.A-9303\\
        NGC 5128 & 3.6 & 0.0018 & S0 pec$^{[2]}$ & 8.18 & -* & 1.2 - 1.7 & 094.B-0298\\

		\hline
	\end{tabular}
\end{table*}

\section{Stellar population analysis}
\label{sect:stellar_pops}
\subsection{Stellar kinematics}

We utilized the Galaxy IFU Spectroscopy Tool \citep[GIST,][]{Bittner2019} to calculate the properties of the stellar component from absorption spectroscopy. This pipeline provides a framework for the use of spectral analysis tools on data cubes, with a default setup specific to MUSE. As a first step, we masked background objects and foreground stars. Then, we used a SNR target value of 150 to spatially bin the pixels using the Voronoi tesselation routine \citep{Cappellari2003}. Despite MUSE providing us a larger spectral range, for this part of the analysis, we restricted the spectra to $4750-6000$\,\AA, since beyond $6000$\,\AA\ strong emission and sky lines contaminate both the SNR estimation required for the binning and the measurement of any properties.
We also manually masked sky lines and strong emission lines. 

Through GIST, we ran pPXF \citep{Cappellari2004,Cappellari2017} to measure the kinematics of the underlying stellar population. We used templates from the MILES spectral library \citep{Vazdekis2010}, covering ages between $0.03$ and $14\,$Gyr, and metallicities between $-2.27 < [\textrm{M}/\textrm{H}] < +0.40$. The SSP models used to generate these libraries are based on BaSTI isochrones \citep{Pietrinferni2004,Pietrinferni2006,Pietrinferni2009,Pietrinferni2013} and  the revised Kroupa initial mass function \citep{Kroupa2001}\footnote{We are aware that  for the most massive ones ETGs, the IMF could be dwarf-richer. However, below $<6000$\,\AA, the contribution of IMF variations is almost negligible.}. We used 25 additive Legendre polynomial, and extracted 4 kinematic moments, corresponding to the coefficients in the velocity profile expansion to a Gauss-Hermite series \citep{1993ApJ...407..525V, 1993MNRAS.265..213G}. 

\subsection{Stellar population properties}

\begin{figure}
    \centering
    \includegraphics[width=\linewidth]{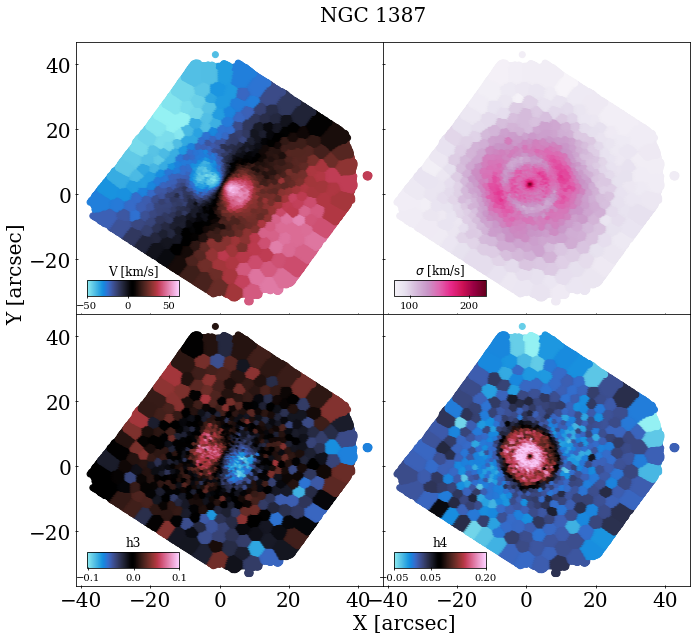}
    \caption{Kinematic properties of the stellar component of NGC\,1387 as obtained with GIST.}
    \label{fig:1387kin}
\end{figure}

In Figure\,\ref{fig:1387kin} we show the line-of-sight (LOS) velocity $V$, the dispersion, $\sigma$, and the parameters $h_{3}$ and $h_{4}$, for NGC\,1387 \footnote{Equivalent figures for the other nine galaxies in the sample are presented in Appendix \ref{sec:app1}}.  

The results are consistent with those by \cite{Iodice2019}, where the same data are also analyzed with pPXF, but outside of the GIST framework. The bar in the centre of NGC\,1387 defines the structure we see in the kinematics, generating the high $\sigma$ values ($\sigma>200$\,km/s) in the central region and probably influencing the annular structure seen in the $\sigma$ distribution as well. $h_{3}$ and $h_{4}$ are close to 0 as expected, though in the central region, the latter goes to higher values, which is consistent with the presence of the bar, disk and bulge in this area.

\begin{figure}
    \centering
    \includegraphics[width=\linewidth]{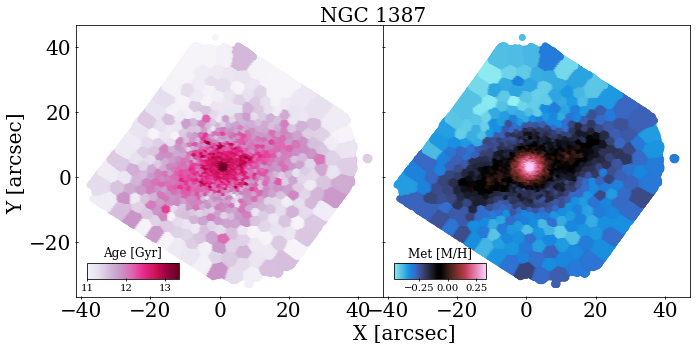}
    \caption{Age and metallicity of the stellar component of NGC\,1387 as obtained with GIST.}
    \label{fig:1387ssp}
\end{figure}

We then determined the age and metallicity distribution of the stellar population of the galaxies in the sample by using pPXF in GIST to calculate a non-parametric star formation history, using 25 multiplicative Legendre polynomials and a regularisation error of 1 \citep{Cappellari2017}. In Figure\,\ref{fig:1387ssp} we show the results for NGC\,1387, for which we measure old ages, consistent with an ETG, and a steep metallicity gradient, going from super-solar to sub-solar, at increasing galactocentric distances. In the region where the bar is situated, we find younger, more metal-rich stars, indicating a connection between the structure and a starburst episode. 

For the rest of the sample, we compared with ATLAS3D \citep{Cappellari2011}, Fornax3D \citep{Martin-Navarro2021} and \cite{Guerou2016}, to ensure overall compatibility between our results and those previously available. Some slight differences were found, however, average properties were found to be consistent.

\section{Detecting planetary nebulae}
\label{sect:PN_detection} 
\subsection{Detecting and measuring the [\ion{O}{iii}] $5007 \AA$ emission line} 

As mentioned above, the main issue we face when it comes to detecting emission lines from unresolved PNe in the inner region of galaxies is the intensity of the stellar continuum. \rvw{We followed the approach described by \cite{Roth2021}, creating a continuum image by co-adding layers from the cube redward of the redshifted [\ion{O}{iii}] $5007 \AA$ line.}
%we modeled the continuum locally, fitting a fifth order polynomial to the spectra in each spaxel, generating, in this way, a continuum cube. We subtracted this cube from the original one, which results in the emission cube that we then searched for PNe with the DELF method \citep{Roth2021}.
%used in the analysis that follows.

The DELF method makes use of the high spectral resolution in MUSE data cubes to extract images with a very narrow spectral range, which reduces \rvw{systematic} noise. Splitting the narrow-band filter in this manner increases the signal-to-noise ratio (SNR) of the emission lines, facilitating the detection of fainter objects even in the central regions.

\rvw{We extracted fifteen slices with the shortest spectral range possible given the spectral resolution available ($1.25$\,\AA), from which we subtract the scaled continuum image to obtain \textit{diff} images}. These slices are centered in the redshifted position of the [\ion{O}{iii}] $5007 \AA$ line, and they cover the expected range of radial velocities for PNe in the centres of ETGs, $\pm 500$\,km/s. Considering the typical spectral FWHM in MUSE data, $\sim 2.6$\,\AA, we expect emission lines to take up at least three slices. To enhance the detection of faint objects, we built narrow band images by adding up to three consecutive slices, thus increasing the flux in the emission line. Using \textsc{SExtractor} \citep{Bertin1996}, we detected the point-like sources present in each of these images, considering as positive detections those that appear in at least three adjacent images (see also \citealt{Roth2021}). Since there is still a significant amount of noise in these images, we used the variance maps as weight images to avoid spurious detections. 

\begin{figure}
    \centering
    \includegraphics[width=\linewidth]{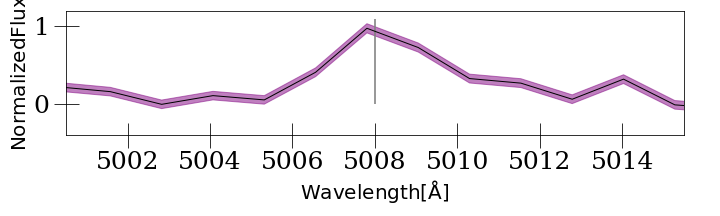}
    \caption{Zoom-in on the region of the [\ion{O}{iii}] line using the spectra constructed from the aperture photometry method described in the text. A grey vertical line indicates the restframe wavelength in air of the [\ion{O}{iii}] line.}
    \label{fig:spec}
\end{figure}

We obtained the aperture magnitude for each of our detected sources in each slice, using an aperture of 5\,pixels and creating a short spectrum as shown in the top panel of  Figure\,\ref{fig:spec}. We fitted a Gaussian profile to this short spectrum as a way to automatically reject any remaining spurious detections. These are mainly sources where there is extended emission or where an issue in the continuum generates a false detection, such as groups of pixels in the edge of the field with arbitrarily negative and positive values. In these short spectra, the continuum subtraction presents some issues that generate noise in the calculation of the flux. In order to obtain more accurate magnitudes, we obtained the aperture photometry of every single slice throughout the entire spectra of each source, using the original cube. We defined  an annulus around each source when performing aperture photometry from which the sky is measured locally. Subtracting this value from the flux measured in the source resulted in a cleaner continuum subtraction. Through this process, we computed a measurement of the flux in every slice of the spectrum, which allows to obtain the total flux of the line, $F_{5007}$, by adding the flux measured in each slice. \rvw{We apply an aperture correction to these flux measurements.} We then calculated the magnitude of the line following the Jacoby relation (1989):

\begin{equation}
\label{eq:magnitude}
    m_{5007} = -2.5\,\rm{log}F_{5007}-13.74.
\end{equation}

\begin{figure*}
    \centering
    \includegraphics[width=\linewidth]{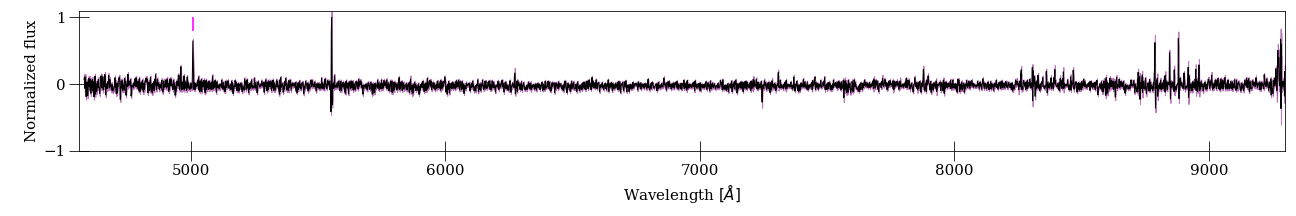}
    \caption{Full MUSE spectrum for one of the sources classified as PNe in NGC\,1387. A pink vertical line indicates the position of the [OIII line].}
    \label{fig:spectr_full}
\end{figure*}

The magnitudes obtained in this way are compatible with the ones resulting from the integration of the Gaussian profile fitted to the short spectrum, but the errors are smaller using aperture photometry, and this method is much more efficient when applied to faint objects than the integration of the line. The spectrum obtained in this manner is show in Figures\,\ref{fig:spectr_full} and \ref{fig:spec} for the case of NGC\,1387.


% \section{Identifying PN candidates}

\subsection{Detecting interlopers} 

The next step is to identify interlopers  (e.g. other sources that can be detected in the [\ion{O}{iii}] $5007 \AA$ filters) in our catalog by first measuring the flux of other emission lines and then applying cuts based on established emission-line ratios. In order to do this, we extracted the spectra corresponding to each detection using aperture photometry throughout the entire continuum-subtracted cube. This allowed us to identify emission lines typically used in diagnostic diagrams, and to measure their flux using the same method as described for [\ion{O}{iii}]. The most common interlopers are mainly \ion{H}{ii} regions and supernovae remnants (SNR). 

\new{Ly$\alpha$ emitters and other background galaxies at high redshift may also show up in narrow-band surveys because their lines can be redshifted into the bandpass. However, background galaxies are easily distinguishable from PNe in MUSE data since they present multiple emission-lines, and the Ly$\alpha$ line has a characteristic profile through which it can be identified}. The [\ion{O}{iii}] line is a doublet (4959\AA, 5007\AA), which can also be used to confirm the emission that is being detected, though the second line is fainter and therefore harder to detect. 

\ion{H}{ii} regions and PNe both emit H$\alpha$ and [\ion{O}{iii}] lines, but the ratio \mar{[\ion{O}{iii}]/H$\alpha$} tends to differ. In \ion{H}{ii} regions, the H$\alpha$ emission is usually more intense than emission in the [\ion{O}{iii}] doublet \citep{Shaver1983}, while the opposite happens in PNe spectra \citep{Baldwin1981}. In particular, \cite{Ciardullo2002} found that the typical ratios of emission lines in PNe can be described with the following equation: 

\begin{equation}
4 > \log{ \frac{\rm [\ion{O}{iii}]}{\rm H{\alpha}\,+[\ion{N}{ii}]}  } > -0.37\,\textrm{M}_{\rm [\ion{O}{iii}]} - 1.16
\label{eq:ciard}
\end{equation}

\begin{figure}
    \centering
    \includegraphics[width=\linewidth]{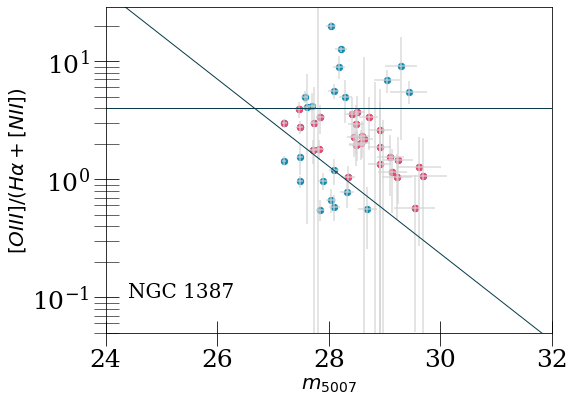}
    \caption{$\rm{[\ion{O}{iii}]}/(H{\alpha}+\rm{[\ion{N}{ii}]})$ line ratios as a function of [\ion{O}{iii}] apparent magnitude for the entire sample of objects detected in NGC\,1387 using the [\ion{O}{iii}] field. The solid lines indicate the criteria which separates PNe candidates (pink dots that fall between the lines) from contaminants (blue dots above the horizontal line or below the diagonal) following \protect\cite{Ciardullo2002}.}
    \label{fig:diagnp}
\end{figure}

In Figure\,\ref{fig:diagnp} we show our diagnostic diagram for NGC\,1387, with the relations from Equation\,\ref{eq:ciard} plotted with solid lines. The diagnostic diagrams for the entire sample can be found in Appendix \ref{sec:app2}. The applied criterion is empirical, and though there are others based on spectroscopy, we consider it is enough for our sample based on \cite{Kreckel2017}. 

\new{Notably, the region of the diagram where the ratio of emission lines is $\log{ \frac{\rm [\ion{O}{iii}]}{\rm H{\alpha}\,[\ion{N}{ii}]}  }>4$  shows several sources for many of our galaxies. Due to measurement errors, some of the objects in this region could be genuine PNe, however, after careful inspection of the spectra, we cannot say this with certainty for all objects above the line. We therefore only consider the objects within the regions defined by Equation\,\ref{eq:ciard} as bona-fide PNe and note that the total number of PNe could be underestimated by a factor of 1.1-1.7.}

\begin{figure}
    \centering
    \includegraphics[width=\linewidth]{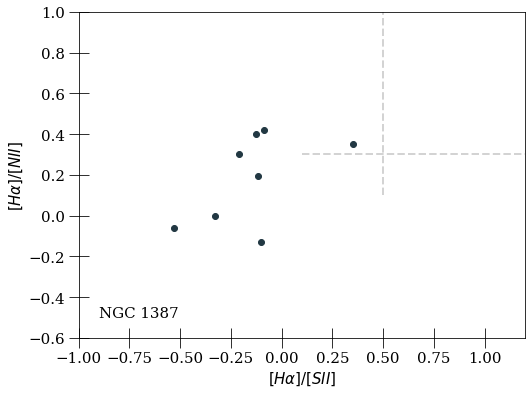}
    \caption{$\rm{H}\alpha/\rm{[\ion{N}{ii}]}$ as a function of $\rm{H}\alpha/\rm{[\ion{S}{ii}]}$ for all sources detected in the [\ion{O}{iii}] filter in NGC\,1387. Black dots indicate sources likely to be SNR. The dashed lines separate the area where \ion{H}{ii} regions would be found, following the criteria from \protect\cite{Sabbadin1977}.}
    \label{fig:diag2np}
\end{figure}

For sources considered interlopers or contaminants, we can analyse other line ratios in order to discriminate \ion{H}{ii} regions from \mar{supernovae remnants (SNRs)}. For objects in which [\ion{N}{ii}] and [\ion{S}{ii}] lines are strong enough, we follow the analysis proposed originally by \cite{Sabbadin1977}. In Figure\,\ref{fig:diag2np} we show the distribution of the interlopers found in the NGC\,1387 sample with strong [\ion{N}{ii}] and [\ion{S}{ii}] lines. All of these objects have line ratios consistent with SNRs. Considering we are looking at the central region of an early-type galaxy, it is not unexpected that we find no \ion{H}{ii} regions, which are more common in LTGs with younger populations and ongoing star formation. 
Our catalog for NGC\,1387, which started with 63 candidates detected via the DELF technique, is then composed of 26\,PNe, with a magnitude range between $29.8<\mathrm{m}_{5007}<27.5$.

\subsection{Completeness test}
\label{sect:completeness}

Estimating the completeness is necessary for the derivation of the PNLF, and also for the analysis of the PN number density profile. Since we are attempting to recover faint objects from a crowded area with a strong continuum, it is indispensable to obtain accurate completeness levels, both in terms of magnitude and of \new{isophotal} galactocentric distance. 

We built 20 catalogues of $\sim 100$ sources each for which we create a position and a magnitude considering the size and shape of each FoV and a magnitude range of $m_{*}-m_\mathrm{lim}=3$, where $m_{*}$ is the bright cut-off estimated from the SBF distance for each galaxy. We used tasks from \textsc{photutils.datasets} to add the objects to our [\ion{O}{iii}] image where we performed the detections, using an elliptical Moffat profile with standard parameters simulating the PSF of MUSE to shape the sources.

\begin{figure}
    \centering
    \includegraphics[width=\linewidth]{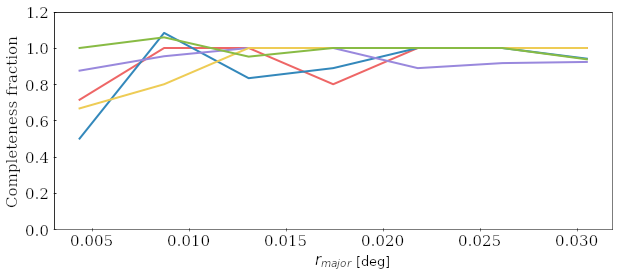}
    \caption{Completeness fraction of several slices in NGC\,1387 in relation to the distance to the galactic centre.}
    \label{fig:compnp1}
\end{figure}

\begin{figure}
    \centering
    \includegraphics[width=\linewidth]{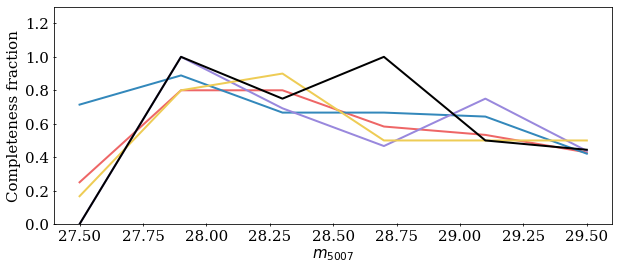}
    \caption{Completeness fraction of several slices in NGC\,1387 in relation to the magnitude of their [\ion{O}{iii}] line.}
    \label{fig:compnp2}
\end{figure}

We then repeated the detection process previously described for PNe candidates and compared the fraction of recovery of sources in relation to the \new{isophotal} galactocentric distance and the [\ion{O}{iii}] line magnitude. For our pilot case, NGC\,1387, we performed this process for each slice separately. Since the trends are comparable across all slices, showing no dependence with wavelength, we performed the process based on a single slice for the remaining galaxies in the sample. In Figures\,\ref{fig:compnp1} and \ref{fig:compnp2}, we show the completeness fractions corresponding to several slices in NGC\,1387 in relation to distance and magnitude. The inner regions present the lowest values of completeness, which is expected due to the brightness of the stellar component increasing towards the centre. In terms of magnitude, the completeness fraction decreases towards the faint end.

\section{Planetary nebulae analysis}
\label{sect:PN_analysis} 
\subsection{Spatial distribution and kinematics}

\begin{figure}
    \centering
    \includegraphics[width=\linewidth]{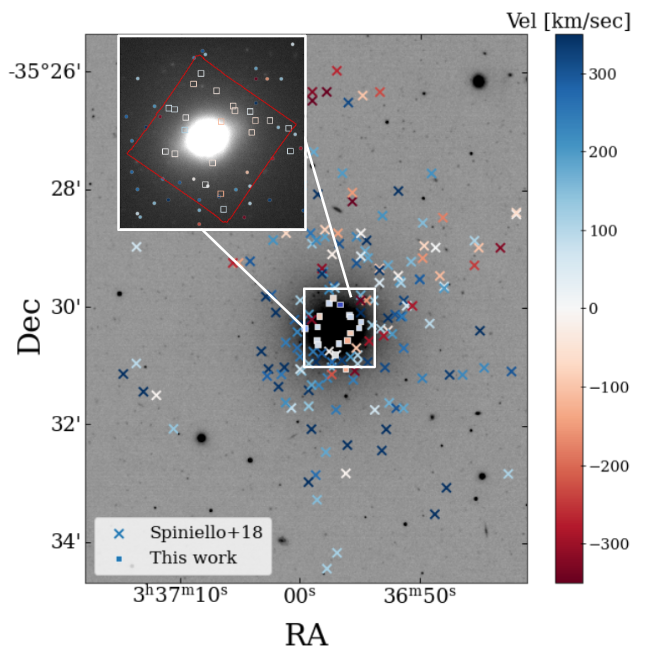}
    \caption{PNe detected in NGC\,1387 overlaid on a DSS2 image of the galaxy. Crosses represent the sample by \protect\cite{Spiniello2018}, and squares, our detections, filling the previous blind spot. The zoomed-in square on the top left highlights our sample, with squares for our sample and circles for the previous catalogue. The red square marks the shape of the MUSE datacube.  The colourbar represents the kinematics of the PNe.}
    \label{fig:pneifu}
\end{figure}

In Figure\,\ref{fig:pneifu}, we show the PNe catalog obtained for NGC\,1387 by \cite{Spiniello2018}, combined with our PNe detections. \cite{Spiniello2018} obtained their sample from VLT-FORS2 observations, using the inverse-dispersion method \citep{1999MNRAS.307..190D}. The observed MUSE field is situated in the central region, where the classical method presents a clear blind spot. In the case of NGC\,1387, this $1'\times\,1'$ region is approximately $5\times5\,\textrm{kpc}^{2}$. There are seven sources in \cite{Spiniello2018} that fall in the edges of our field, and are therefore not detected. Four others fall inside the field, near the edges, and are too weak for our completeness limit. One source is common to both catalogues. We do not incorporate the missing sources to our analysis for consistency since we cannot do this with the rest of the sample.

From the [\ion{O}{iii}] $5007 \AA$ line shift, we obtain the velocity for each PNe, thus obtaining the velocity distribution shown in Figure \ref{fig:pneifu}. The PN velocities are consistent with the velocity dispersion and systemic velocity derived from absorption spectroscopy. Future work will link the stellar and PN kinematics with those at large radii as traced by \cite{Spiniello2018}.

\subsection{Luminosity function}

The PNLF quantifies the number of PNe in terms of the magnitude of the [\ion{O}{iii}] $5007$\,\AA\, line, which is obtained following Equation\,\ref{eq:magnitude}. The analytic fit to the PNLF was first determined for the bulge of M31 \citep{Ciardullo1989}, and has ever since proven to have a near universal shape following Eq. \ref{eq:pnlf}.

% \begin{equation}
%     N(M_{5007})=c_{1}e^{c_{2}M_{5007}}(1-e^{3(M^{*}-M_{5007})}),
% \end{equation}

% where $c_{1}$ is a normalisation constant, and $M^{*}$ is the absolute magnitude of the bright cut-off, which has empirically been shown to be nearly-constant at $M^{*}=-4.51^{+0.02}_{0.04}$ \citep{Ciardullo2002,Ciardullo2004}. The shape of the PNLF is motivated by the physics of PNe, with the first term describing how the surface brightness of the envelope around the central star is fading as it expands with constant velocity \citep{Henize1963}. The slope $c_2 = 0.307$, considered constant in the original implementation of the function, has since been found to vary with the stellar populations. \cite{Longobardi2013} was the first to  introduce a PNLF with variable faint-end slope, which is thought to be connected to the properties of the underlying stellar populations. Steeper PNLFs have been associated with older, metal-poor stellar populations, while flatter slopes may be connected to younger stars \citep{Longobardi2013,Ciardullo2004,Ciardullo2010,Hartke2020}. 

\begin{figure}
    \centering
    \includegraphics[width=\linewidth]{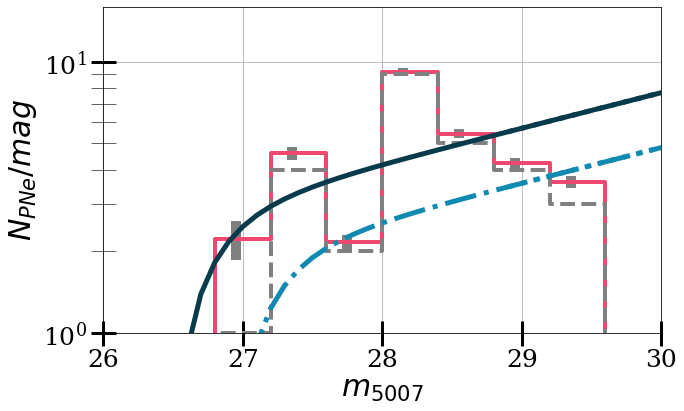}
    \caption{The grey, dashed histogram shows the observed PNLF for NGC\,1387, and in pink solid lines we show the version corrected by completeness, with grey thick bars indicating the errors. The dash-dotted light blue curve shows the best-fit analytic PNLF with a fixed bright cut-off based on the galaxy's distance, and the solid dark blue line indicates the function resulting from fitting the bright-cut off to our data.}
    \label{fig:1387-pnlf}
\end{figure}

Due to the small size of the FoV and the fact that our observed PNLFs are only complete to $1-2$ magnitudes from the bright cut-off $M^{*}$, we refrain from analysing the variation of the slope in this work, and instead use the ``traditional'' \cite{Ciardullo1989} PNLF with $c_2 = 0.307$. The main purpose of the PNLF in this analysis is to extrapolate our sample to a homogeneous limit in the faint end. As shown in Figure\,\ref{fig:1387-pnlf}, we fit two versions of the function, one with a variable bright cut-off and one with this value fixed according to the SBF distance to the galaxy, to the completeness-corrected observed PNLF. This is done for comparison purposes, but to facilitate cross-galaxy comparison, we perform the following extrapolations using the function with the fixed cut-off, fitting only the normalisation $c_1$.
\mar{From Figure~\ref{fig:1387-pnlf}, we can evaluate already that the sensitivity reached by the MUSE data is $\Delta m\sim 2-2.5$ below the PNLF bright cut off.}

\subsection{Surface brightness profile of galaxies and PNe number density}

As PNe trace the stellar populations of ETGs, their number density profile is expected to be proportional to the surface brightness profile of elliptical \citep{Coccato2009,Longobardi2013} and lenticular galaxies \citep{Cortesi2013,Cortesi2013b}. 
% In our case, both profiles are limited in size, but the focus of this analysis is to evaluate the connection between the two in the inner region as to look for potential variations in relation to galactocentric distance.

% In the case of NGC\,1387, \cite{Napolitano2022} combine the PNe catalog obtained in \cite{Spiniello2018} with the GC catalog from \cite{Chaturvedi2022}, and analyse the number density of both tracers in relation to the centre of the galaxy. In their work, the inner region ($R_{gal}=84''$) is excluded. \cite{Napolitano2022} find that PNe expand to a galactocentric radius of approximately $25'$, while GCs reach almost $50'$. They also point out that PNe within $5'$ of the centre of the galaxy are few, and their velocity dispersion is smaller than in the rest of the galaxy. However, this could be caused by the halo of NGC\,1399 which extends to this region and increases the noise. 

For galaxies for which we have data with the FoV of a single MUSE cube, we generate a wide band image from our data cubes in the Cousins $V$ filter using the MUSE pipeline. We then use tasks from \textsc{statmorph} \citep{Rodriguez-Gomez2019} to fit a S\'ersic profile \citep{Sersic1968} to the brightness distribution of the galaxy. 
% In Figure\,\ref{fig:bricomp} we present the resulting surface brightness profile for NGC\,1387. The field of view (FoV) of MUSE is small, so in order to ensure that our results are reasonable, we compare our fit to the one obtained by \cite{Spavone2020} for the inner and outer region, using VLT-OmegaCam. This instrument has a FoV of $1\deg$, which is significantly larger. Our fitted profile falls in the area between their inner and outer region fits, matching the distribution well in the region where PNe are analysed in this work. We consider then that the method is good enough for the purpose of this analysis. 
For the closest galaxies in our sample, where we have a mosaic of several cubes, sometimes with different sensitivities, we use surface brightness profiles obtained from the literature (all references listed in Table\,\ref{tab:surflit}), and fit S\'ersic profiles to them.

The number density profile of our PNe samples is obtained by splitting the area into concentric regions, and dividing the amount of PNe detected in each of these regions by its area. The PNe density profile is defined as:

\begin{equation}
\mu_{\text{PN}}(r) = -2.5\,\log_{10}\Sigma_{\text{PN}}(r)+\mu_{0},
\end{equation}

where $\mu_{0}$ is a constant that we fit to relate the observed PN number density with the stellar surface brightness, and $\Sigma_\mathrm{PN}$ is the PNe number density, corrected by spatial completeness and limited in magnitude by the brightness completeness. 
% In Figure\,\ref{fig:pbri_off} we show the comparison between both profiles.

% \begin{figure}
%     \centering
%     \includegraphics[width=\linewidth]{plots-new/surfbright-preoff.png}
%     \caption{S\'ersic fit to the surface brightness profile for NGC\,1387 in the $V$-band shown in black solid line. Pink crosses indicate the PNe density, with the errorbars signaling the size of the bin in the x-axis and the density error in the y-axis.}
%     \label{fig:pbri_off}
% \end{figure}

\begin{figure}
    \centering
    \includegraphics[width=\linewidth]{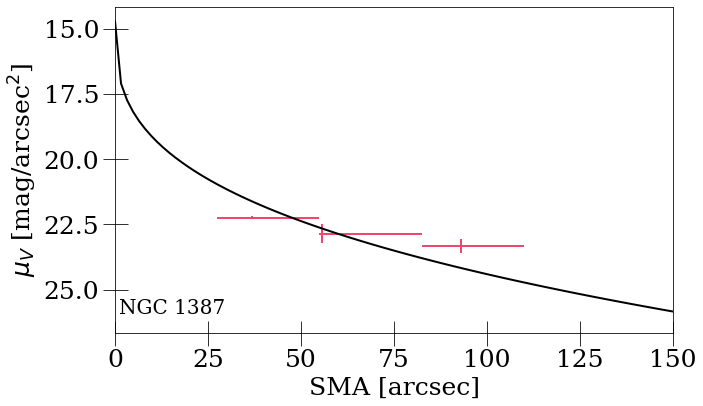}
    \caption{S\'ersic fit to the surface brightness profile for NGC\,1387 in the $V$-band shown in black solid line. Pink crosses indicate the PNe density offset by the fitted constant $\mu_{0}$, with the errorbars indicating the size of the bin in the x-axis and the density error in the y-axis.}
    \label{fig:pbri}
\end{figure}

As the $\alpha$-parameter relates the number of PNe with the total luminosity of a galaxy, we first need a bolometric correction, which, following B06, is defined as:

\begin{equation}
I = 10^{-0.4(BC_{V}-BC_{\odot})}10^{-0.4(\mu-K)},
\end{equation}

$BC_{\odot}$ is the solar bolometric correction and $BC_{V}$ is the bolometric correction for a stellar population model using the $V$ filter, which can be assumed to have a fixed value of $-0.85$ with 10\% accuracy (B06). $K$ is the conversion factor from the units of surface brightness in the chosen band to the physical units $L_{\odot}\textrm{pc}^{-2}$, defined as $K=M_{\odot,\text{filter}}+21.572$\footnote{\url{https://web.archive.org/web/20070718015507/http:/www.ucolick.org/~cnaw/sun.html}}. We then calculate the $\alpha$-parameter using the following equation:

\begin{equation}
\alpha=\frac{10^{-0.4(\mu_{0}-K-(BC_{\odot}-BC_{V}))}}{s^{2}},
\end{equation}

where $s=D/206265$ is a scale factor related to the distance to the galaxy, $D$. For NGC\,1387, we obtain $\mu_{0}=18.14\pm0.4\,\rm{mag}$ and $\alpha=(3.25\pm1.88)\times10^{-8}\,\rm PN\,L_{BOL}^{-1}$. Figure\,\ref{fig:pbri} shows the result of offsetting the PNe density profile by this value of $\mu_{0}$. From this fit, we can calculate the $\alpha$-parameter. \rvw{Although we perform a correction for the area not covered by the annuli in which we measure the PNe density, we exclude annuli for which either the completeness or the covered area is less than 50\% when we perform the fit to measure the offset. Points excluded are shown as black dots in the plots or are excluded from the range displayed.} We consider the $\alpha$-parameter as $\alpha_{2.5}$, meaning our magnitudes are between the bright cut-off, $m_{*}$, and $m_{*}+2.5$, \mar{as shown previously in Figure~\ref{fig:1387-pnlf}}. This is done for homogeneity, since the completeness levels vary through the sample, and to facilitate future comparisons.

Since completeness limits vary for each galaxy, and in order to obtain a value more directly comparable to the literature, we consider a magnitude range of $\Delta m=2.5\,\rm{mag}$ from the bright cut-off. In the case of NGC\,1387, our completeness limit is $m=29$, which is half a magnitude fainter. From the PNLF we can obtain the difference between these magnitude ranges, following

\begin{equation}
\centering
    \Delta m_{c} = \frac{\int^{m^{*}+2.5}_{m^{*}}N(m)\mathrm{d}m}{\int^{m_{c}}_{m^{*}}N(m)\mathrm{d}m}
\end{equation}

This allows us to extrapolate and to obtain $\alpha_{2.5}$, with $\alpha_{2.5}=\Delta m_{c} \times \alpha_{m_{c}}$. This results in $\alpha_{2.5}=(4.32\pm2.49)\times10^{-8}\,\rm PN\,L_{BOL}^{-1}$ for NGC~1387. A summary of the $\alpha$-parameters for all galaxies in the sample, as well as the results from the stellar population analysis, can be found in \mar{Table~\ref{tab:PNresults}.}

\begin{table}
    \centering
    \begin{tabular}{|c|c|}
        \hline
        Galaxy & Reference \\
        \hline
       \new{NGC 1387} & \cite{Li2011} \\
       \new{NGC 1404} & \cite{Li2011} \\
       NGC 1399 & \cite{Caon1994} \\
       NGC 3115  & \cite{Capaccioli1987} \\
       \new{NGC 3379} & \cite{Watkins2014}, \cite{Lee2016}, \cite{Hartke2020} \\
       \new{NGC 4365} & \cite{Kormendy2009} \\
       \new{NGC 4374} & \cite{Kormendy2009} \\
       NGC 4472 & \cite{Kormendy2009} \\
       NGC 4594 & \cite{Li2011} \\ 
       NGC 5128 & \cite{Li2011} \\
       
         \hline
    \end{tabular}
    \caption{Literature references for galaxies whose surface brightness profile we did not measure ourselves}
    \label{tab:surflit}
\end{table}

\section{Discussion}
\label{sect:discussion}
\subsection{Comparison with existing catalogues}
\rvw{Three of the galaxies in our sample, NGC\,1399, NGC\,1404 and NGC\,4472, were previously studied in \cite{Jacoby2024}, using the DELF method but performing the detection of sources visually, blinking through the slices. These three particular cubes each have their own challenges. The version available in the archive of the reduced mosaic for NGC\,1399, as shown in \cite{Jacoby2024}, presents an imperfect flat-field corrections and the noise varies significantly across the galaxy. In the case of NGC\,1404, there are two available cubes, which we analyse invidually and then combine, whereas \cite{Jacoby2024} reduces the data from scratch. Finally, NGC\,4472 has the shortest exposure times in both samples, resulting in very shallow cubes. For the latter cases, we recover approximately half the sources present in the previous work, and some new ones. For NGC\,1399, we find 9 PNe, in stark contrast to the over 100 found in \cite{Jacoby2024}. It is expected that we should find less sources with an automated detection mechanism, and it is particularly challenging for the spatially variable background for emission lines, as mentioned by \cite{Soemitro2023}. In our case, the initial detection finds up to 70\% of the sources in the previous catalogues. Interestingly, most of them do not meet our criteria when fitting a Gaussian to the emission line, and some are not considered PNe when evaluating the diagnostic diagrams.}

\subsection{The luminosity-specific PN number as tracer of stellar population properties}

% The basis of the original analysis done in B06 regarding the $\alpha$-parameter is based on Figure\,10, which shows its variation in relation to the colour of the host galaxy both in observations and in models. Two main things are noted in terms of early-type galaxies:  redder ellipticals show poorer PN populations per unit bolometric luminosity; and despite subtle trends, the PN rate per unit galaxy luminosity remains fairly constant across all morphologies. The first has been shown to remain true when considering different components of a single galaxy, e.g. in M87 \citep{Longobardi2013}, where the inner red halo has a lower $\alpha$ value that increases in the outer, bluer region.  This is seen as well in M49 \citep{Hartke2017}, in M31 \citep{Bhattacharya2019} and in M105 \citep{Hartke2020}, where the $\alpha$-parameter is shown to be tracing distinct populations across the galaxy.

The starting point of the original analysis done in B06 regarding the $\alpha$-parameter is Figure\,10 therein, which shows its variation in relation to the colour of the host galaxy both in observations and in models. LTG on the blue end have, on average, higher \new{observed} $\alpha$-parameter \new{value}s than red ellipticals, which, however, present a larger scatter. This relation with colour has been shown to remain true when considering different components of a single galaxy, e.g. in M87 \citep{Longobardi2013}, where the inner red halo has a lower $\alpha$ value that increases in the outer, bluer region.  This is seen as well in M49 \citep{Hartke2017}, in M31 \citep{Bhattacharya2019} and in M105 \citep{Hartke2020}, where the $\alpha$-parameter is shown to be tracing distinct populations across the entire radial extent of a galaxy.

\begin{table*}
\begin{tabular}{|l|c|c|c|c|c|}
\hline
Galaxy & Age [Gyr] & [Fe/H] & $\sigma$ [km/s] & $\alpha \,[\times10^9]$ & $\alpha_{\rm err} \,[\times10^9]$\\
\hline
\quad NGC 1387 & 11.5 $\pm$ 0.4 & -0.29 $\pm$ 0.09 & 102.62 $\pm$ 20.23 & $11.66$ & $11.71$ \\
\quad NGC 1399 & 12.3 $\pm$ 0.7 & -0.11 $\pm$ 0.29 & 229.88 $\pm$ 10.51 & $1.47$ & $1.37$ \\
\quad NGC 1404 & 12.7 $\pm$ 0.3 & -0.08 $\pm$ 0.10 & 189.87 $\pm$ 18.96 & $4.82$ & $2.14$ \\
\quad NGC 3115 & 10.9 $\pm$ 1.3 & -0.08 $\pm$ 0.17 & 123.95 $\pm$ 26.76 & $4.55$ & $1.55$ \\
\quad NGC 3379 & 12.7 $\pm$ 0.6 & -0.12 $\pm$ 0.14 & 159.05 $\pm$ 30.79 & $3.22$ & $0.98$ \\
\quad NGC 4365 & 11.4 $\pm$ 0.9 & 0.14 $\pm$ 0.09 & 233.39 $\pm$ 10.38 & $2.12$ & $0.37$ \\
\quad NGC 4374 & 12.7 $\pm$ 0.2 & 0.00 $\pm$ 0.07 & 276.97 $\pm$ 6.76 & $1.10$ & $1.19$ \\
\quad NGC 4472 & 9.1 $\pm$ 1.4 & -0.44 $\pm$ 0.05 & 257.26 $\pm$ 62.54 & $4.19$ & $8.24$ \\
\quad NGC 4594 & 8.3 $\pm$ 2.5 & -0.11 $\pm$ 0.22 & 196.84 $\pm$ 60.08 & $6.36$ & $1.66$ \\
\quad NGC 5128 & 4.6 $\pm$ 1.4 & -0.77 $\pm$ 0.52 & 191.73 $\pm$ 92.78 & $7.08$ & $1.07$ \\
\hline
\end{tabular}
\caption{Mean values of the stellar population properties obtained from GIST, and $\alpha$-parameter up to 2.5 magnitudes from the bright cut-off of the PNLF for the entire sample.}\label{tab:PNresults}
\end{table*}

\begin{figure}
    \centering
    \includegraphics[width=\linewidth]{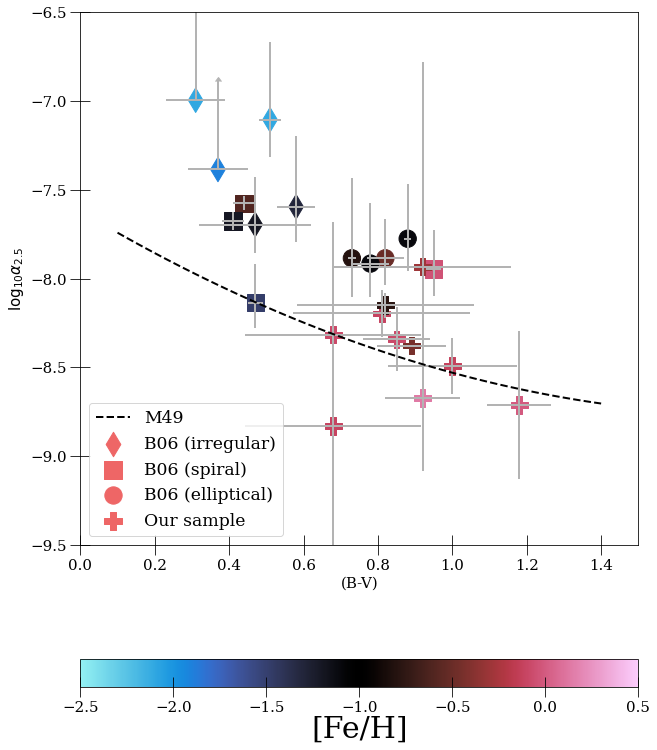}
    \caption{Luminosity-specific PN number $\alpha_{2.5}$ versus the (B-V) colour of the host galaxy, coloured by the metallicity of the galaxy. Thick crosses indicate our sample. The dashed line indicates the results for M49 obtained by \protect\cite{Hartke2017}.}
    \label{fig:albv-met}
\end{figure}

\begin{figure}
    \centering
    \includegraphics[width=\linewidth]{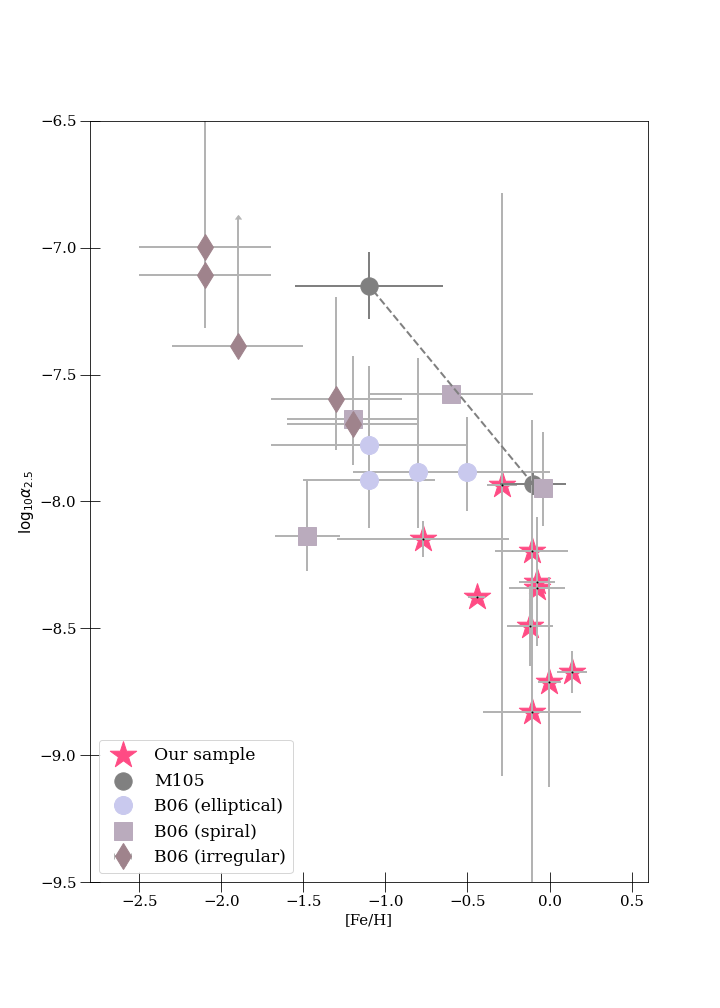}
    \caption{Luminosity-specific PN number $\alpha_{2.5}$ versus the metallicity of the galaxy for the B06 sample in pastel colours. Pink stars show the results from our sample. Grey dots joined by a dashed line show the results from \protect\cite{Hartke2022} for M105.}
    \label{fig:almet}
\end{figure}

\begin{figure}
    \centering
    \includegraphics[width=\linewidth]{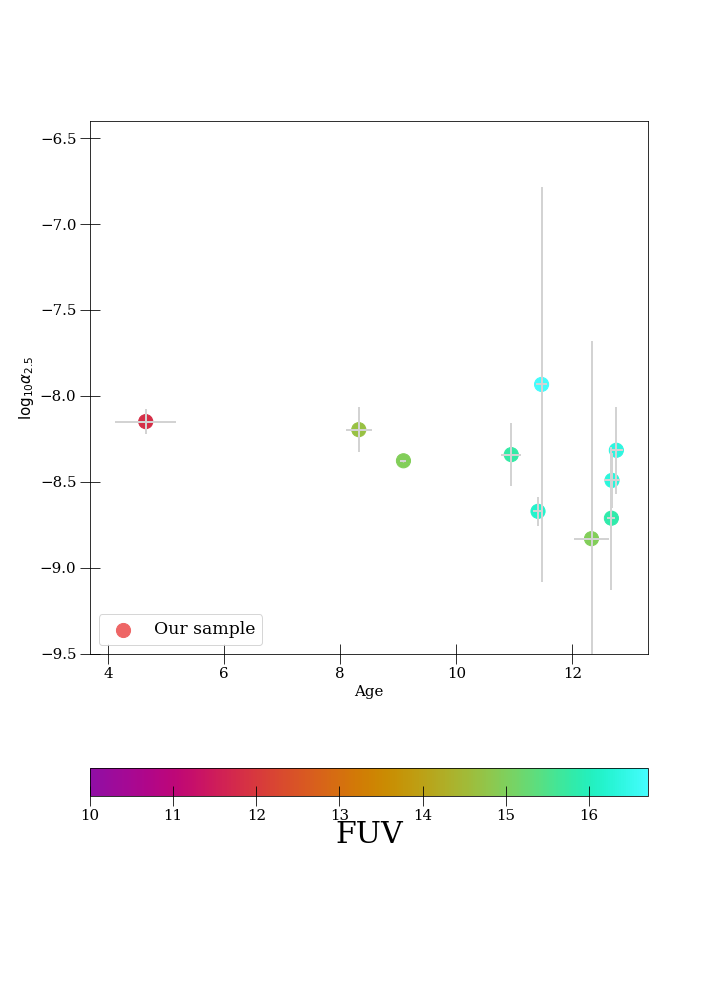}
    \caption{\rvw{Luminosity-specific PN number $\alpha_{2.5}$ versus the age of the host galaxy for the galaxies in our sample.}}
    \label{fig:almet2}
\end{figure}

Figure\,\ref{fig:albv-met} shows the relation between $\alpha_{2.5}$ and the $(B-V)$ colour of the host galaxy for the B06 galaxies and our sample, as obtained \new{from \cite{deVaucouleurs1991} for NGC\,3379, NGC\,4365, NGC\,4374 and NGC\,4472, and from \cite{Ho2011} for the rest of the sample, both corrected for extinction}. The colour bar indicates the mean of the metallicity of the stellar population. \rvw{Our sample of galaxies follows the relation, although with some scatter. In Figure\,\ref{fig:almet}, we show the relation between $\alpha_{2.5}$ and the mean metallicity of the corresponding data cube for its host galaxy. Most galaxies in the sample follow the relation traced by M105, minus the previously described outliers that fall well below it. From Figure\,\ref{fig:almet2}, which shows the relation between $\alpha_{2.5}$ and the age of the host galaxy, it seems clear that metallicity is driving the correlation seen in colour. Recent theoretical results from \cite{Valenzuela2025} find a strong correlation of $\alpha_{2.5}$ with metallicity and a weaker one with age, although \mar{similarly to previous results in B06}, the theoretical correlation \mar{would predict an} exact opposite trend \mar{to the one obtained } observationally, with higher \new{predicted} values of $\alpha_{2.5}$ in galaxies with higher metallicities.}
%The large scatter in $\alpha_{2.5}$, especially for red and metal-rich ETGs, blurs the relation, although the bluer end still presents larger values overall when compared to the redder galaxies. Figure\,\ref{fig:almet} shows the relation between $\alpha_{2.5}$ and the metallicity of the host galaxy. Although the scatter remains, and the metallicity does not vary significantly across galaxy centres, our sample mostly spans lower values of $\alpha_{2.5}$ than galaxies with larger metallicities. Comparing galaxies within the sample does not show any significant trend, albeit the metallicity range is small. Complementary, Figure\,\ref{fig:almet2} shows the same relation but coloured by the mean age of the stellar population for our sample only in order to provide a more in-depth look at the age-metallicity degeneracy. The relation appears to be significatively less dependent on age than on metallicity, but the scatter is too large to draw any conclusions.

It is noteworthy that the relation between the $\alpha$-parameter and colour or metallicity are much more prominent in the studies of individual galaxies, i.e. M49 and M105. In both of these cases, the comparison is performed between distinct halo populations with less weaker age and metallicity gradients than the stellar cores studied in this work. 

\begin{figure}
    \centering
    \includegraphics[width=\linewidth]{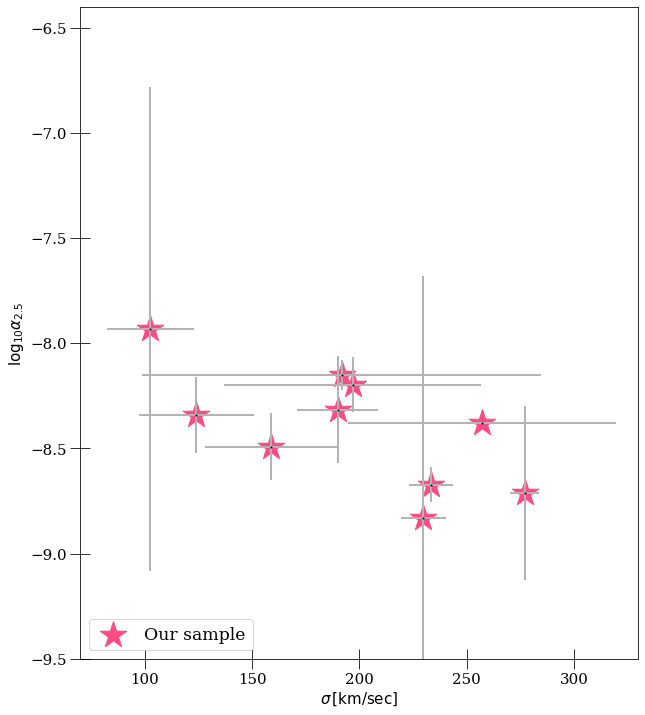}
    \caption{Luminosity-specific PN number $\alpha_{2.5}$ versus the velocity dispersion of the host galaxy for the galaxies in our sample.}
    \label{fig:aldisp}
\end{figure}

Finally, Figure\,\ref{fig:aldisp} shows the relation between $\alpha_{2.5}$ and the velocity dispersion as measured at the centre of the galaxy. It can be argued that the relation shown in B06 where higher velocity dispersions correspond to lower number densities prevails in this sample. This in turn may be interpreted as a relation between $\alpha_{2.5}$ and the total mass of the galaxy.

\subsection{PNe visibility lifetime}

\begin{figure}
    \centering
    \includegraphics[width=\linewidth]{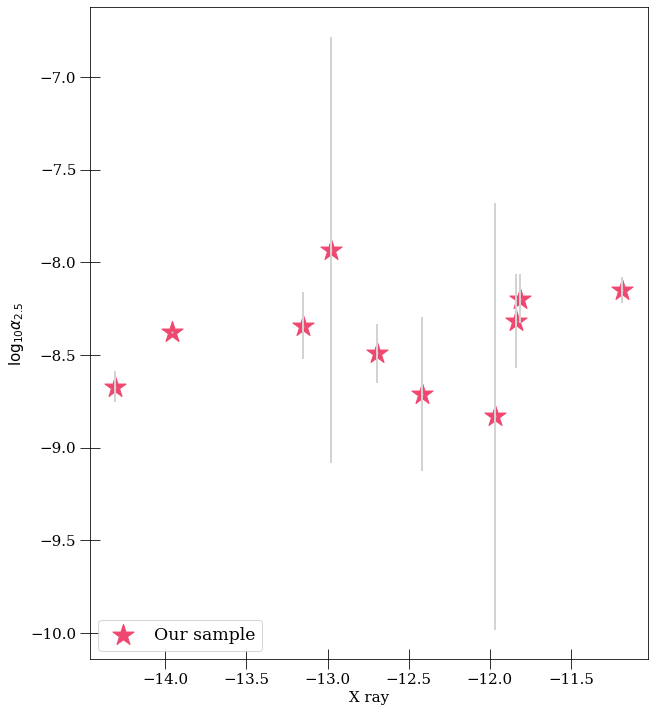}
    \caption{Luminosity-specific PN number $\alpha_{2.5}$ versus the X-ray magnitude of the host galaxy for the galaxies in our sample.}
    \label{fig:alxray}
\end{figure}

The $\alpha$-parameter is defined as $\alpha=\textrm{N}_{\textrm{PN}}/\textrm{L}_{\textrm{bol}}$, where the number of observable PNe, $\textrm{N}_{\textrm{PN}}$, depends on the evolutionary flux $\mathcal{B}$ of  and the visibility lifetime $\tau_{\textrm{PN}}$ according to the SSP theory developed by \cite{Renzini1986} and \cite{Buzzoni1989}. A lower value of $\alpha$ may then be a result of shorter PN lifetimes, which may be due to evolutionary reasons linked to the properties of the stellar populations, or to the interaction of the interstellar medium with the PN envelopes. In the central regions of massive ETGs, hot gas in high densities could cause ram-pressure stripping of the envelopes, effectively reducing $\tau_{\textrm{PN}}$ \citep{Dopita2000,Martin2002,Villaver2005}. However, in Figure\,\ref{fig:alxray} no clear correlation is seen between the intensity of the X-ray emission in the central region of the galaxies and the $\alpha$ parameter within the current sample. The X-ray values used in this analysis are reported in Table\,\ref{tab:fot}.

\begin{figure}
    \centering
    \includegraphics[width=\linewidth]{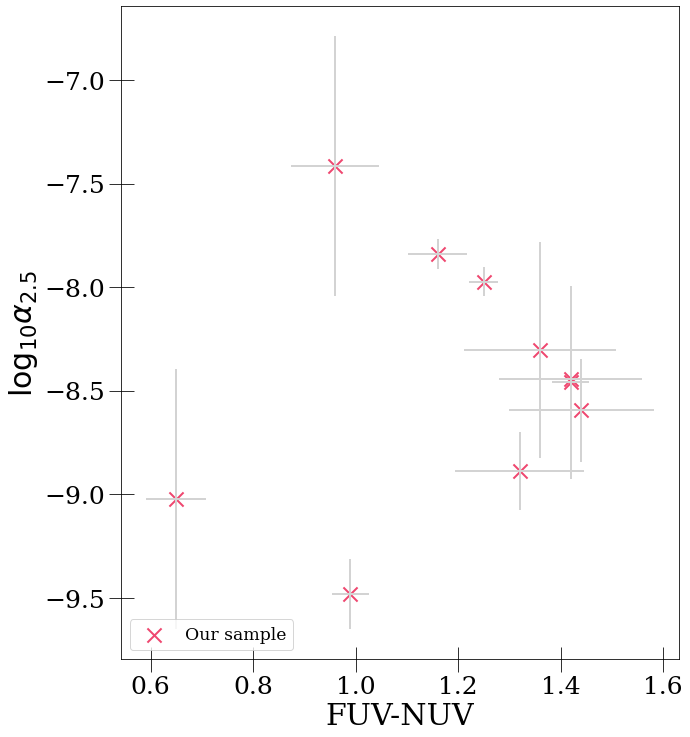}
    \caption{Luminosity-specific PN number $\alpha_{2.5}$ versus FUV-NUV colour of the host galaxy for the galaxies in our sample.}
    \label{fig:aluv}
\end{figure}

In a different wavelength range, the UV excess has been shown to anti-correlate with $\alpha$ (B06), \new{and in more detail in \cite{Ciardullo2005}, using $m_{1500}-V$, which has a dynamic range 6 times larger than FUV-NUV}. The evolutionary pathways of stars are nuanced, and the visibility lifetime of a PNe is very dependent on how hot the core is, and on the asymptotic giant branch transition time. Stars with low-mass cores occupy the hotter horizontal branch, and skip the PNe phase since the envelope fades very quickly. This affects the spectral energy distribution of the galaxy, resulting in UV-enhanced ellipticals being expected to haver lower $\alpha$ values. In Figure\,\ref{fig:aluv} we show that this anti-correlation is present in our sample despite the scatter. An outlier in the  $\rm log\alpha_{2.5}$ vs FUV-NUV plot, NGC\,1399, place in the lower left corner, disrupts the contour of the proposed anti-correlation. NGC\,1399 is a complex galaxy, and in future work we will analyse its $\alpha$-parameter and its potential variations, studying the effects of AGN activity on PNe populations. Expanding our sample beyond ETGs and performing the observations and analysis in a systematic way will provide a stronger conclusion.

\begin{figure*}
    
    \centering
    \includegraphics[scale=0.4]{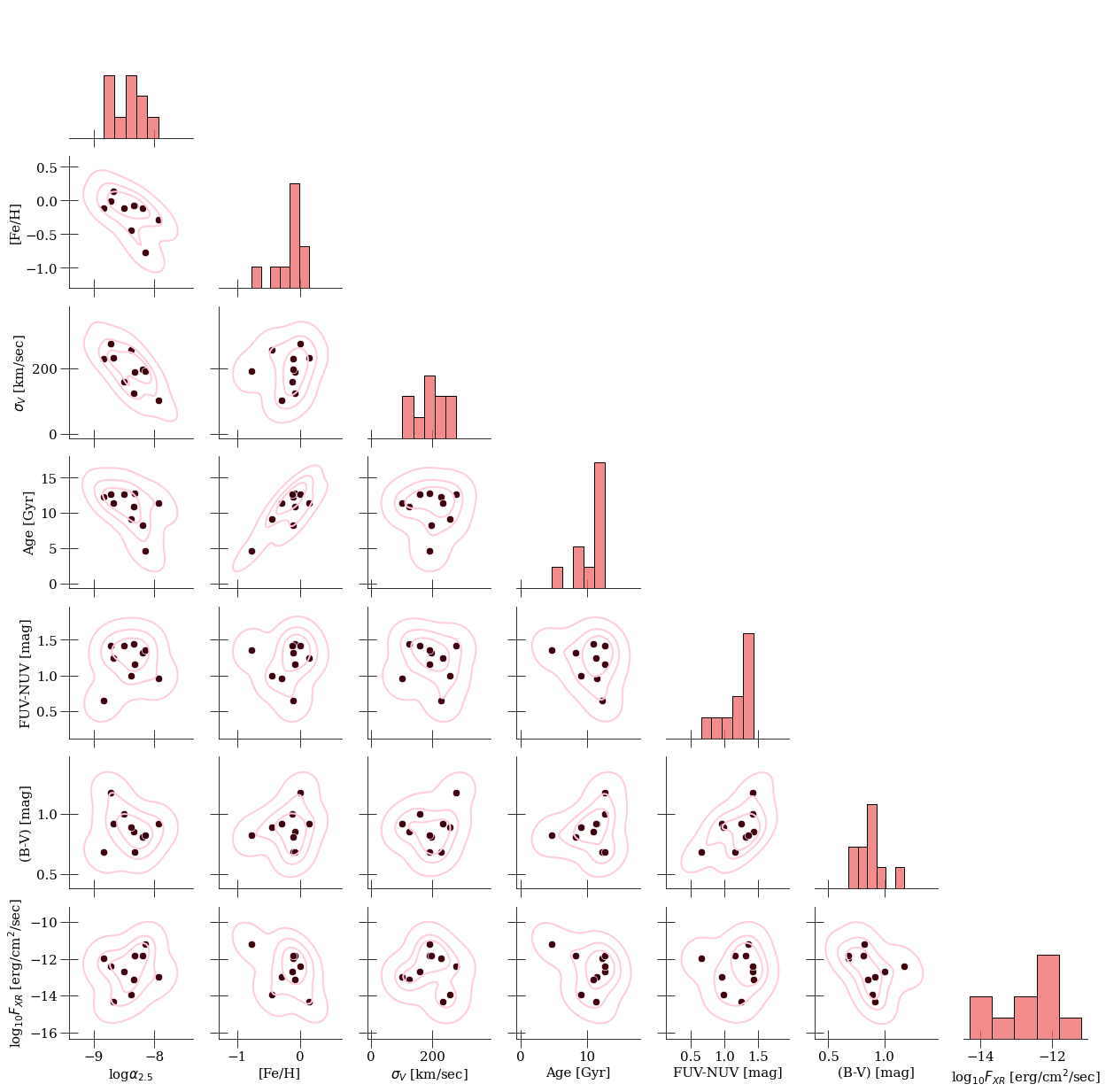}
    \caption{Corner plot showing the relations between all analysed parameters. Measurements are in black points, with solid pink lines indicating density estimations obtained with KDE. The upper plots show histograms corresponding to each parameter.}
\label{fig:cornerplot}
\end{figure*}

Finally, Figure\,\ref{fig:cornerplot} shows a corner plot with all the analysed parameters in order to visualize all relations. From this Figure, we see there \rvw{although there is scatter, there is a correlation between $\alpha_{2.5}$ and metallicity, as well with the FUV excess.}

\begin{table}
    \centering
    \begin{tabular}{c|c|c|c}
    \hline
Galaxy & FUV [mag] & NUV [mag] & $\textrm{log}_{10}(\textrm{X-ray} [\textrm{erg}/\textrm{cm}^2/\textrm{s}])$  \\
\hline
NGC $1387^{1}$ & $16.71\pm0.05$ & $15.75\pm0.07$ & N/A \\
NGC $1399^{1}$ & $14.96\pm0.05$ & $14.31\pm0.03$ & $-11.30\pm0.002$ \\
NGC $1404^{1}$ & $16.37\pm0.05$ & $15.21\pm0.03$ & $-11.54\pm0.002$ \\
NGC $3115^{1}$ & $15.78\pm0.14$ & $14.34\pm0.02$ & $-12.83\pm0.003$ \\
NGC $3379^{1}$ & $16.33\pm0.03$ & $14.91\pm0.02$ & $-13.31\pm0.05$ \\
NGC $4365^{2}$ & $16.18\pm0.02$ & $14.93\pm0.02$ & $-12.97\pm0.002$ \\
NGC $4374^{1}$ & $15.82\pm0.13$ & $14.40\pm0.05$ & $-11.85\pm0.004$ \\
NGC $4472^{1}$ & $14.98\pm0.03$ & $13.99\pm0.02$ & $-11.30\pm0.002$ \\
NGC $4594^{2}$ & $14.68\pm0.11$ & $13.36\pm0.06$ & $-12.60\pm0.003$ \\
NGC $5128^{1}$ & $11.83\pm0.07$ & $10.47\pm0.13$ & N/A \\
\hline
    \end{tabular}
    \caption{UV magnitudes and X-ray fluxes for the central regions of each galaxy obtained from the literature. FUV and NUV were obtained from \protect\cite{Bai2015} for galaxies signaled as $[1]$ and \protect\cite{Zaritsky2015} for $[2]$ , which are corrected for Galactic extinction using the Galactic Reddening Map of \protect \cite{Schlegel1998} and the reddening law of \protect \cite{Cardelli1989}. X-ray magnitudes are from \protect \cite{Babyk2018} except for NGC\,3115 \protect \citep{Norris2006} and NGC\,4594 \protect \citep{Gonzalez-Martin2009}. We found no values in the literature for NGC\,1387 and NGC\,5128.}
    \label{tab:fot}
\end{table}

\subsection{Limitations and prospects of archival data}

Astronomers currently have access to unpredecedented amounts of high quality data, made available through careful upkeeping of observatory archives. Although archival data is a precious tool that allows for works such as this one, providing the possibility to test pipelines and methods on large data sets with varying quality, it comes with limitations due to the mix of observing conditions, mainly caused by the different research questions behind the original proposals. 

As mentioned, IFU spectroscopy enables detecting PNe in regions where previous methods faced insurmountable challenges. However, several considerations must be made even with this powerful tool at hand. Galaxies with active galactic nuclei (AGN), such as NGC\,1316, proved immune to the method, since the emission lines of the AGN were too intense and widespread across the cubes to untangle them from PNe candidates. Galaxies at smaller distances, such as NGC\,5128, occupy a FoV that requires constructing a mosaic from several pointings, which is expensive in telescope time despite the shorter exposure time per pointing and may result in varying depths across the field. Finally, characterizing the age and metallicity of the stellar component is highly challenging in galaxies with dusty centres, as seen in NGC\,5128 as well as NGC\,4594. For future work, an ideal sample should be observed considering the needed sensitivity to reach the $m^*+2.5$ limit in the [\ion{O}{iii}] $5007\AA$ line as to avoid extrapolating, focusing on ETGs between 15 and 25\,Mpc. To apply this method, both the spatial and spectral resolution are highly relevant. The spectral sampling of MUSE, $1.25$\AA, is key for DELF to work, and similar or higher spectral resolutions should be aimed for also in future works, \new{keeping in mind that increasing the spectral resolution too much would resolve the emission lines and hinder the process rather than help it}. Spatially, the method relies on detecting PNe as point-like sources, which implies a constraint on image quality according to the distance of the host galaxy. In the case of MUSE, using its adaptive optics system may significantly boost PN detections \citep[e.g.][]{2023A&A...671A.142S}. Finally, the BlueMUSE instrument \citep{2019arXiv190601657R} with a similar instrument design compared to MUSE, but with a larger FoV and a bluer spectral coverage will also allow direct abundance determinations of extragalactic PNe in the centres of galaxies.  

% \begin{figure*}
%     \centering
%     \includegraphics[width=\linewidth]{plots-new/alpha-bv.png}
%     \caption{Luminosity-specific PN number $\alpha_{2.5}$ versus the (B-V) colour of the host galaxy for the B06 sample in pastel colours. Pink circles with a white filling show the results from our sample.}
%     \label{fig:albv}
% \end{figure*}

% \begin{figure}
%     \centering
%     \includegraphics[width=\linewidth]{plots-new/alpha-met3.png}
%     \caption{Luminosity-specific PN number $\alpha_{2.5}$ versus the metallicity of the host galaxy for the galaxies in our sample, coloured by the FUV emission of the host galaxy.}
%     \label{fig:almet3}
% \end{figure}

% not necessarily to be included in paper but important to remember 
% in order to obtain the zero-point magnitude for the slices, i created a tophat-like filter that has a value of zero everywhere outside of the slice using muse_cube_filter and i got zeropoints in the header
% HIERARCH ESO DRS MUSE FILTER ZPVEGA = 21.032514 / [mag] Vega-mag zeropoint      
% HIERARCH ESO DRS MUSE FILTER ZPAB = 20.156914 / [mag] AB-mag zeropoint         !! these aren't necessarily useful bc they correspond to the entire cube and slices come from the emission-line cube 
\section{Summary and conclusions}
\label{sect:conclusions}
In this work, we analyse the central regions of a sample of ten ETGs, detecting and identifying PNe using MUSE archival data. We focus on the $\alpha$-parameter, i.e. the luminosity-specific number density of PNe, and its potential correlations with a variety of properties of the underlying stellar population. We find that the $\alpha$-parameter remains tightly constrained within a narrow range of values, as shown for galaxy haloes across the literature. \rvw{Our sample of galaxies follows the previously proposed correlation between the $\alpha$-parameter and the metallicity of the host galaxies, while the correlation with age is very weak, implying the correlation with colour shown in B06 is \mar{possibly driven} by the metallicity of the underlying subpopulations.} The FUV excess anti-correlates with the $\alpha$-parameter, validating previous results \mar{ and confirming a tension between theoretical predictions and observed PN population properties, despite inclusion of then most recent advanced PAGB evolution tracks}. A larger sample is required for future work to confirm these results and further \mar{improve} our understanding of the physics behind these relations.

\section*{Acknowledgements}

We thank the referee for an insightful report and helpful comments that improved this work. We thank Martin Roth for providing guidance for using DELF. We thank Felipe Gran for sharing part of the code used to perform aperture photometry on the cubes. We also thank Elizaveta Sazonova for useful discussions and suggestions. This research was supported in part by Perimeter Institute for Theoretical Physics.  Research at Perimeter Institute is supported by the Government of Canada through the  Department of Innovation, Science and Economic Development and by the Province of  Ontario through the Ministry of Research and Innovation. A.E. and J.H acknowledge the financial support from the visitor programme of the Finnish Centre for Astronomy with ESO (FINCA), funded by the Academy of Finland grant nr 306531. This work made use of AstroPy \citep{astropy1,astropy2,astropy3}, Matplotlib \citep{matplotlib}, NumPy \citep{numpy} Pandas \citep{pandas}, PhotUtils \citep{photutils}, SciKit \citep{scikit-learn} and SciPy \citep{scipy}.

%%%%%%%%%%%%%%%%%%%%%%%%%%%%%%%%%%%%%%%%%%%%%%%%%%
\section*{Data Availability}

The processed MUSE data are all publicly available from the ESO Science Archive \mar{(DOI: https://doi.eso.org/10.18727/archive/41)} We plan to release our PN catalogues as advanced data products to ESO.



\bibliographystyle{mnras}
\bibliography{biblio} % if your bibtex file is called example.bib

%\begin{thebibliography}{99}
%\end{thebibliography}

%%%%%%%%%%%%%%%%%%%%%%%%%%%%%%%%%%%%%%%%%%%%%%%%%%
\newpage
%%%%%%%%%%%%%%%%% APPENDICES %%%%%%%%%%%%%%%%%%%%%
\appendix

\section{Stellar population properties}
\label{sec:app1}
\begin{figure}
    \centering
    \includegraphics[width=\linewidth]{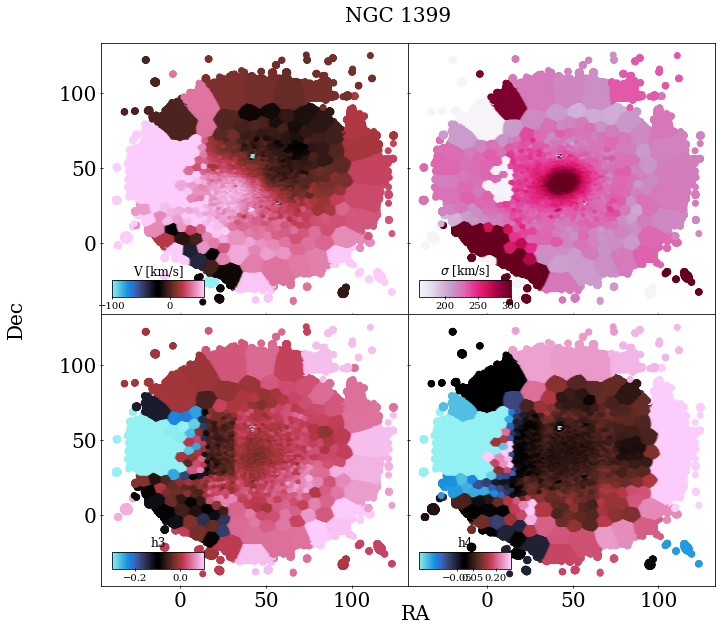}
    \label{fig:1399-kin}
    \caption{Kinematic properties of the stellar component of NGC\,1399 as obtained with GIST.}
\end{figure}

\begin{figure}
    \centering
    \includegraphics[width=\linewidth]{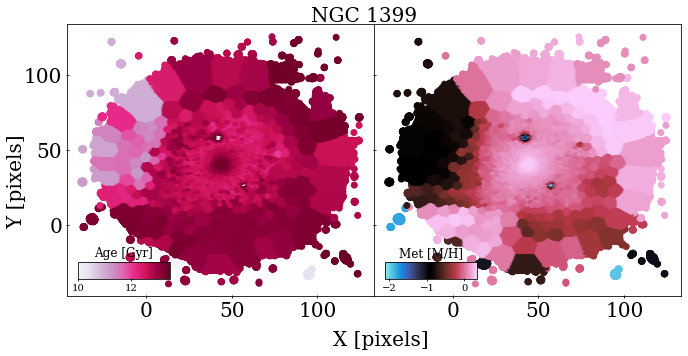}
                \caption{Stellar population properties of the stellar component of NGC\,1399 as obtained with GIST.}
    \label{fig:1399-sfh}
\end{figure}

In this appendix, we present our results from the analysis of the kinematics and properties of the stellar populations in each cube, following the methodology described in Sect.~\ref{sect:stellar_pops}. For each galaxy, we present maps of the stellar kinematics parameterised by $V$, $\sigma$, $h_3$, and $h_4$ (e.g. Fig.~\ref{fig:1399-kin}) and the stellar ages and metallicities (e.g. Fig.~\ref{fig:1399-sfh}. Light-weighted mean values were extracted from these maps, excluding the very central regions where we do not detect PNe (cf. Sect.~\ref{sect:completeness}), to use in the analysis presented in the discussion in Sect.~\ref{sect:discussion}. 

\begin{figure}
    \centering
    \includegraphics[width=\linewidth]{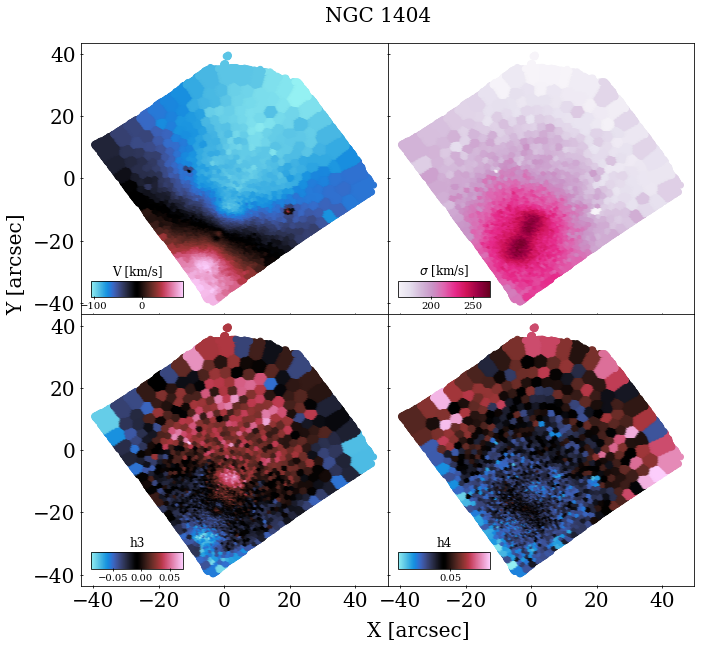}
                \caption{Kinematic properties of the stellar component of NGC\,1404 as obtained with GIST.}
    \label{fig:1404-kin}
\end{figure}

\begin{figure}
    \centering
    \includegraphics[width=\linewidth]{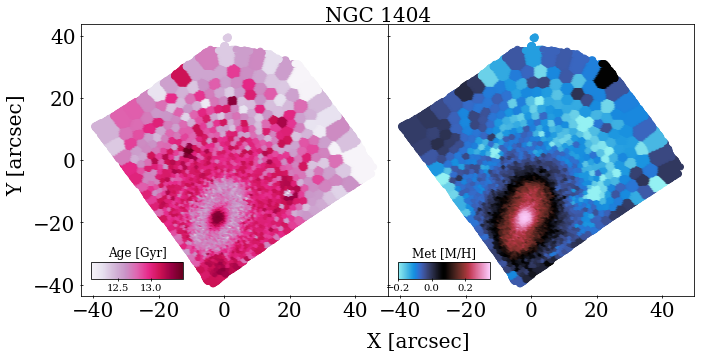}
    \label{fig:1404-sfh}
            \caption{Stellar population properties of the stellar component of NGC\,1404 as obtained with GIST.}
\end{figure}

\begin{figure}
    \centering
    \includegraphics[width=\linewidth]{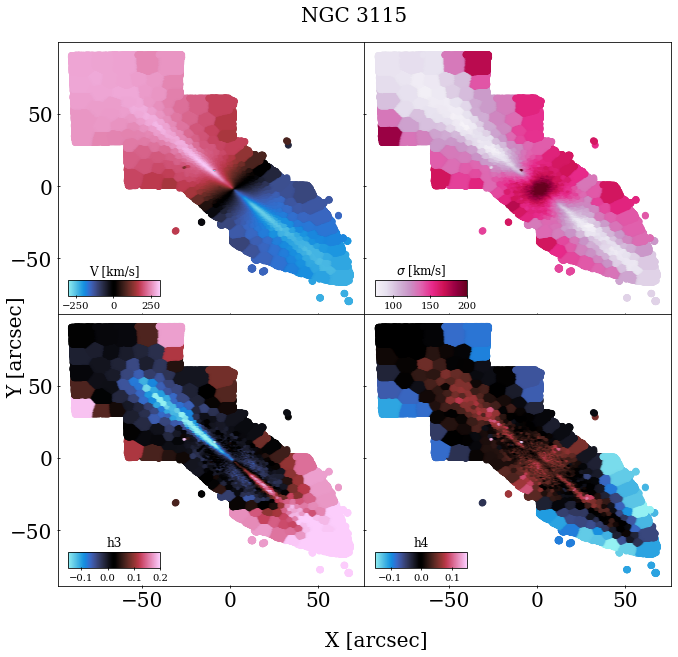}
    \label{fig:3115-kin}
    \caption{Kinematic properties of the stellar component of NGC\,3115 as obtained with GIST.}

\end{figure}

\begin{figure}
    \centering
    \includegraphics[width=\linewidth]{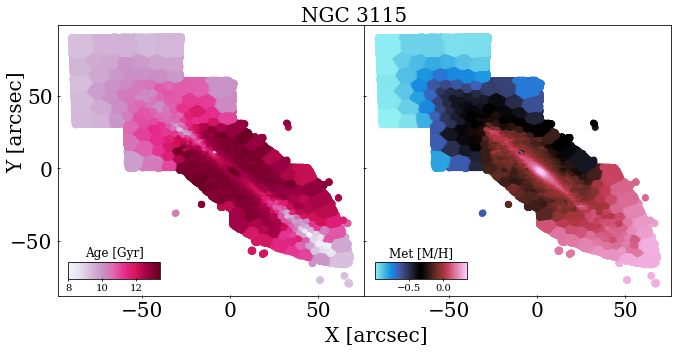}
                \caption{Stellar population properties of the stellar component of NGC\,3115 as obtained with GIST.}
    \label{fig:3115-sfh}
\end{figure}

\begin{figure}
    \centering
    \includegraphics[width=\linewidth]{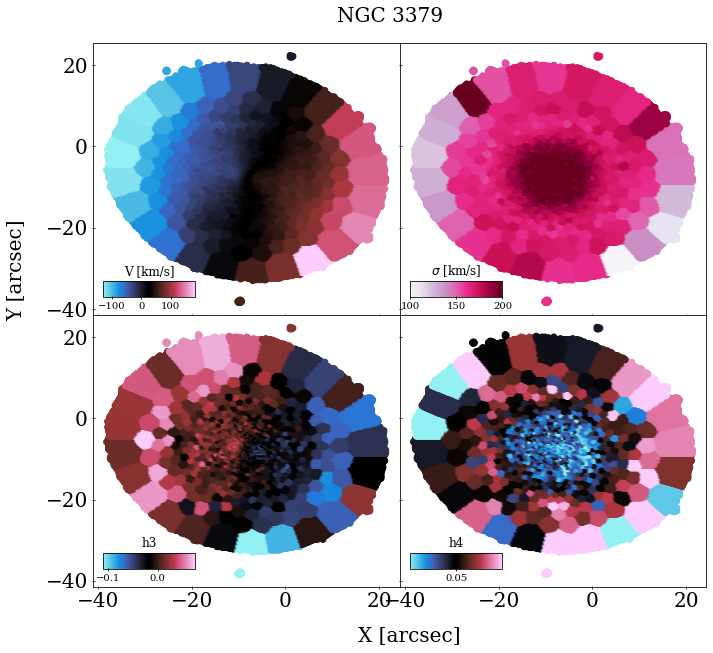}
        \caption{Kinematic properties of the stellar component of NGC\,3379 as obtained with GIST.}

    \label{fig:3379-kin}
\end{figure}

\begin{figure}
    \centering
    \includegraphics[width=\linewidth]{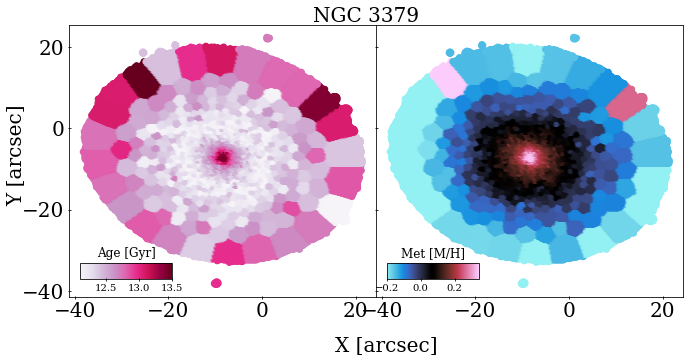}
    \label{fig:3379-sfh}
                \caption{Stellar population properties of the stellar component of NGC\,3379 as obtained with GIST.}
\end{figure}

\begin{figure}
    \centering
    \includegraphics[width=\linewidth]{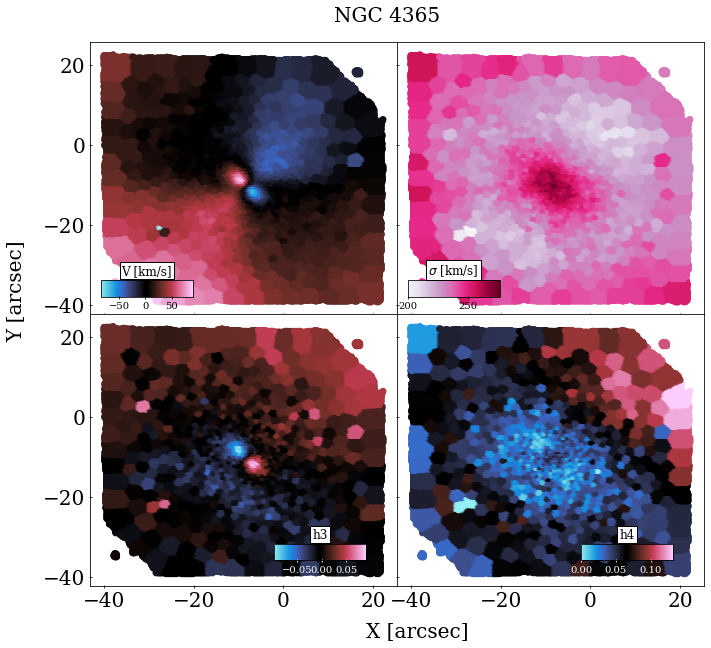}
        \caption{Kinematic properties of the stellar component of NGC\,4365 as obtained with GIST.}

    \label{fig:4365-kin}
\end{figure}

\begin{figure}
    \centering
    \includegraphics[width=\linewidth]{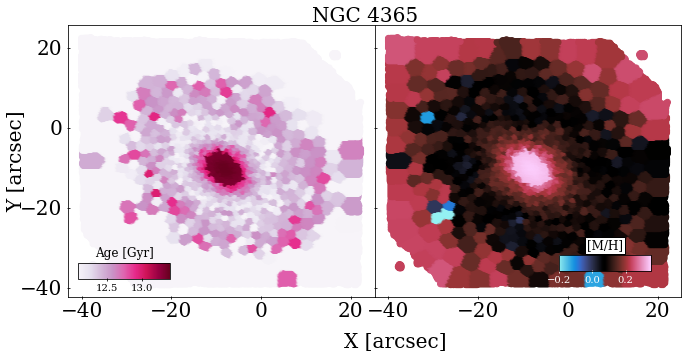}
                \caption{Stellar population properties of the stellar component of NGC\,4365 as obtained with GIST.}
    \label{fig:4365-sfh}
\end{figure}

\begin{figure}
    \centering
    \includegraphics[width=\linewidth]{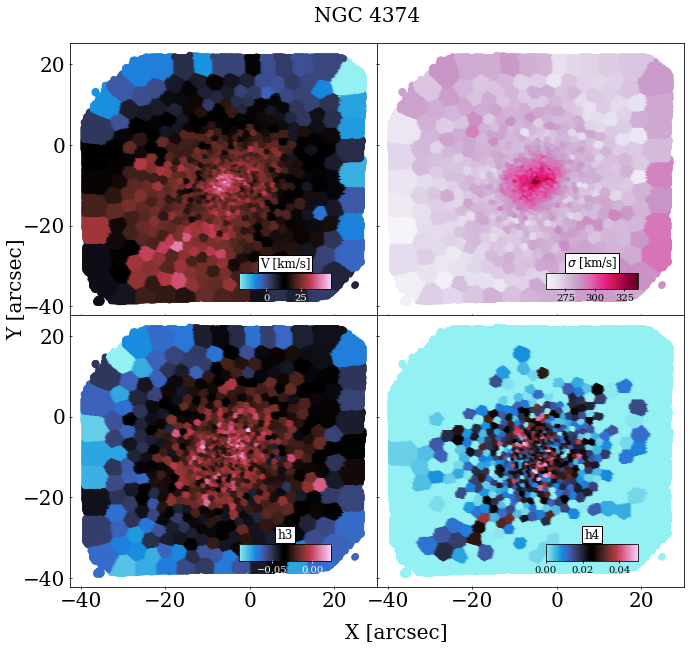}
        \caption{Kinematic properties of the stellar component of NGC\,4374 as obtained with GIST.}

    \label{fig:4374-kin}
\end{figure}

\begin{figure}
    \centering
    \includegraphics[width=\linewidth]{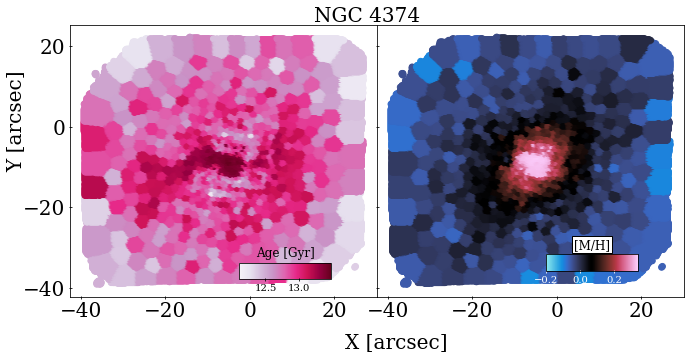}
                \caption{Stellar population properties of the stellar component of NGC\,4374 as obtained with GIST.}
    \label{fig:4374-sfh}
\end{figure}

\begin{figure}
    \centering
    \includegraphics[width=\linewidth]{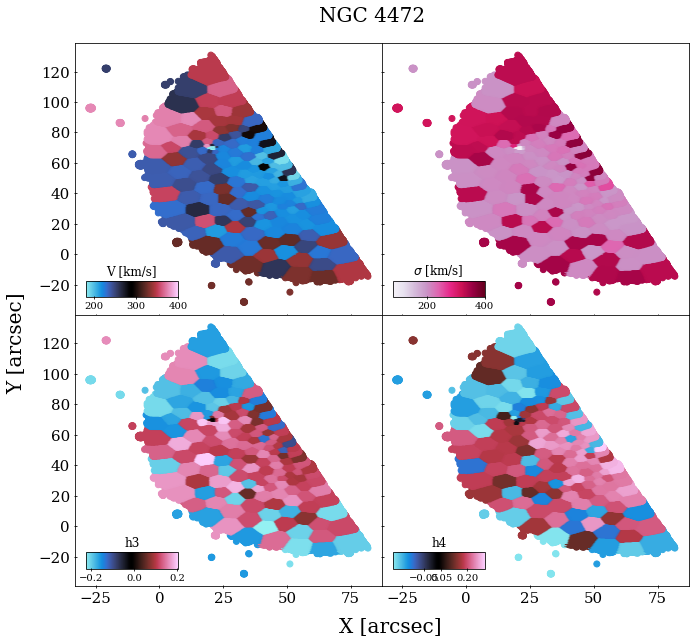}
        \caption{Kinematic properties of the stellar component of NGC\,4472 as obtained with GIST.}

    \label{fig:4472-kin}
\end{figure}

\begin{figure}
    \centering
    \includegraphics[width=\linewidth]{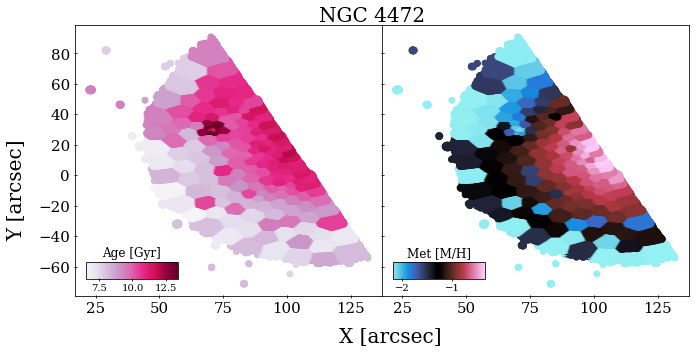}
                \caption{Stellar population properties of the stellar component of NGC\,4472 as obtained with GIST.}
    \label{fig:4472-sfh}
\end{figure}

\begin{figure}
    \centering
    \includegraphics[width=\linewidth]{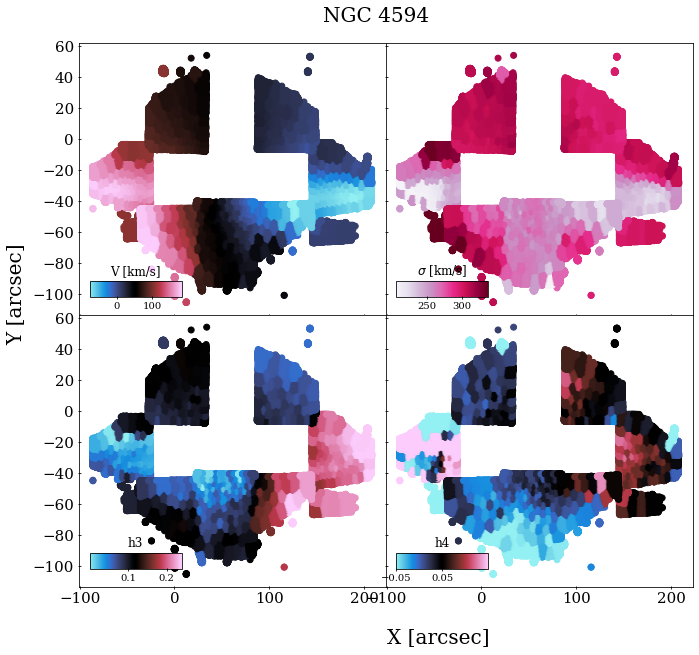}
        \caption{Kinematic properties of the stellar component of NGC\,4594 as obtained with GIST.}

    \label{fig:4594-kin}
\end{figure}

\begin{figure}
    \centering
    \includegraphics[width=\linewidth]{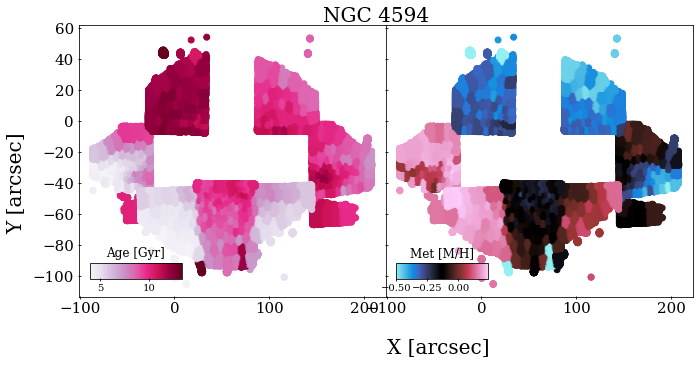}
                \caption{Stellar population properties of the stellar component of NGC\,4594 as obtained with GIST.}
    \label{fig:4594-sfh}
\end{figure}

\begin{figure}
    \centering
    \includegraphics[width=\linewidth]{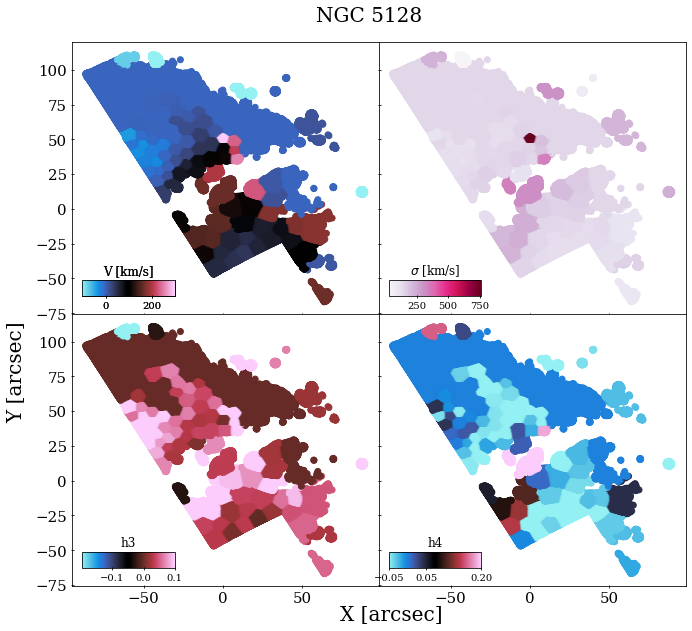}
        \caption{Kinematic properties of the stellar component of NGC\,5128 as obtained with GIST.}

    \label{fig:5128-kin}
\end{figure}

\begin{figure}
    \centering
    \includegraphics[width=\linewidth]{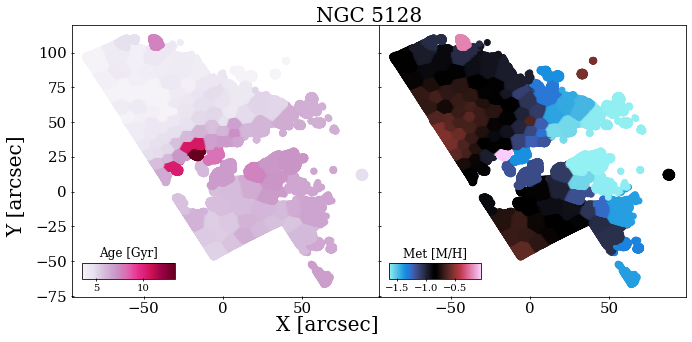}
                \caption{Stellar population properties of the stellar component of NGC\,5128 as obtained with GIST.}
    \label{fig:5128-sfh}
\end{figure}

\begin{figure}
    \centering
    \includegraphics[width=\linewidth]{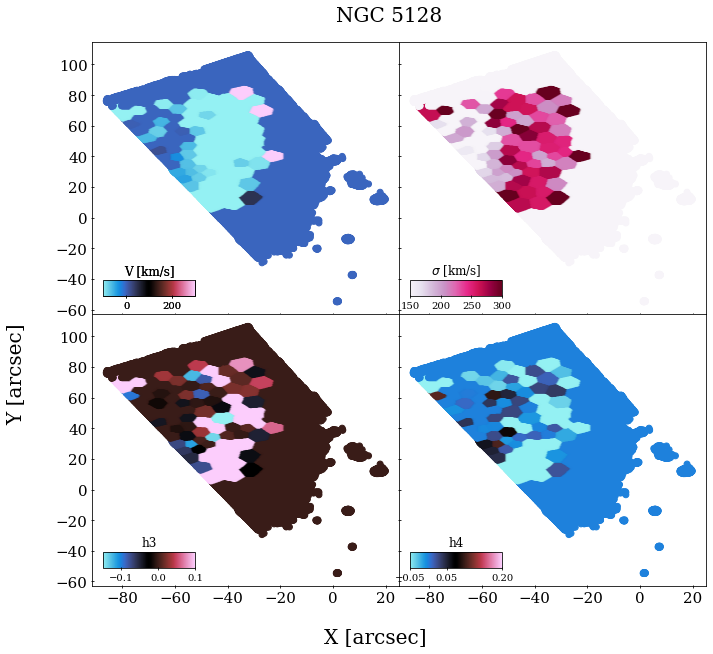}
        \caption{Kinematic properties of the stellar component of NGC\,5128 as obtained with GIST.}

    \label{fig:51282-kin}
\end{figure}

\begin{figure}
    \centering
    \includegraphics[width=\linewidth]{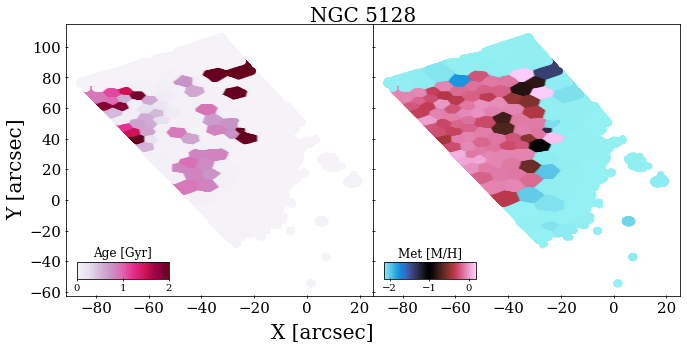}
                \caption{Stellar population properties of the stellar component of NGC\,5128 as obtained with GIST.}
    \label{fig:51282-sfh}
\end{figure}

\begin{figure}
    \centering
    \includegraphics[width=\linewidth]{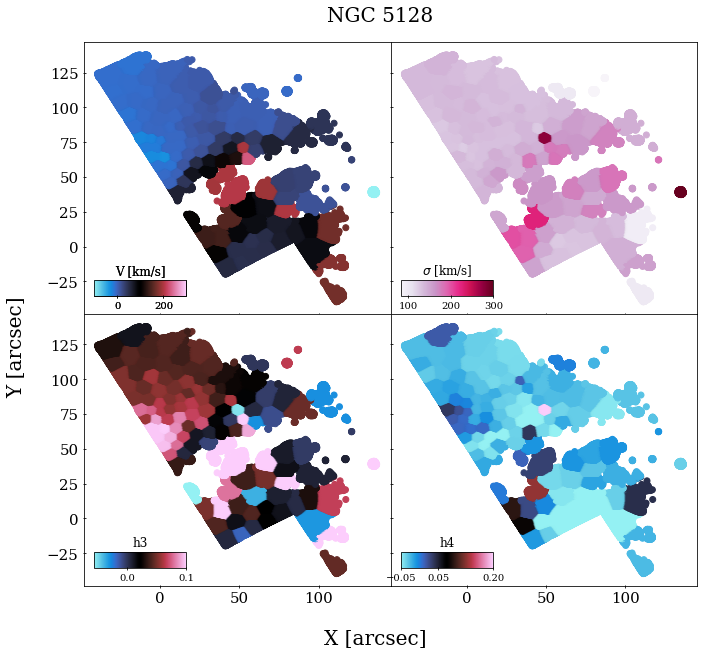}
        \caption{Kinematic properties of the stellar component of NGC\,5128 as obtained with GIST.}

    \label{fig:51283-kin}
\end{figure}

\begin{figure}
    \centering
    \includegraphics[width=\linewidth]{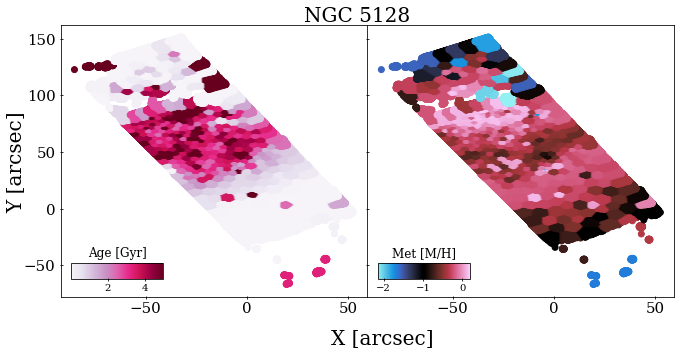}
                \caption{Stellar population properties of the stellar component of NGC\,5128 as obtained with GIST.}
    \label{fig:51283-sfh}
\end{figure}

\begin{figure}
    \centering
    \includegraphics[width=\linewidth]{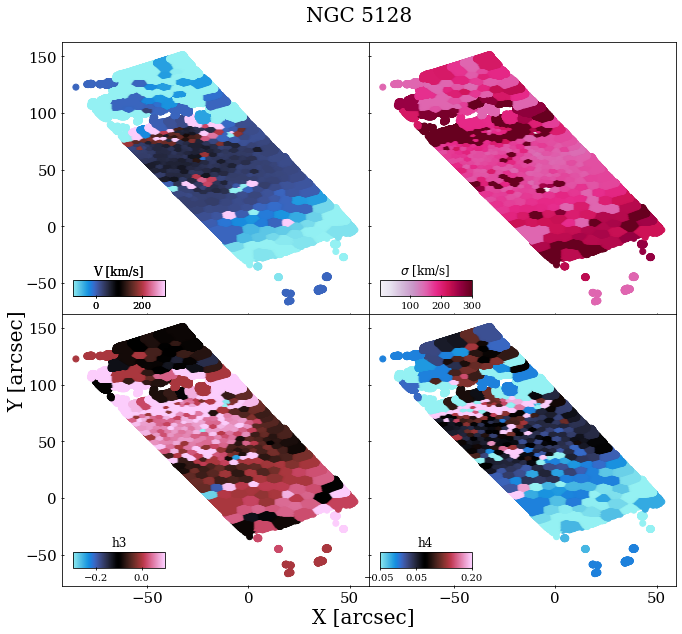}
        \caption{Kinematic properties of the stellar component of NGC\,5128 as obtained with GIST.}

    \label{fig:51284-kin}
\end{figure}

\begin{figure}
    \centering
    \includegraphics[width=\linewidth]{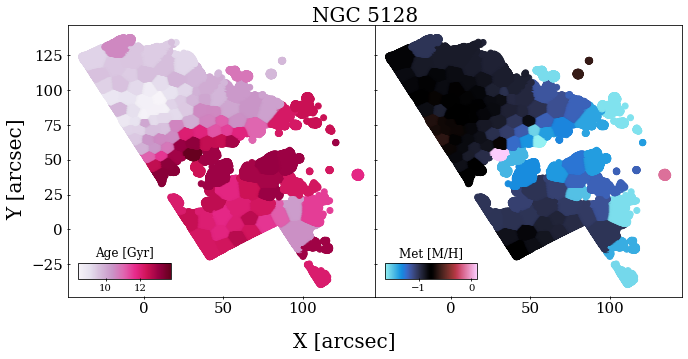}
            \caption{Stellar population properties of the stellar component of NGC\,5128 as obtained with GIST.}

    \label{fig:51284-sfh}
\end{figure}

\section{Planetary nebulae analysis}
\label{sec:app2}
In this Appendix,  we present the diagnostic diagrams for the sources detected in each galaxy, with blue points showing sources considered interlopers, and pink ones, PNe. The solid lines indicate the criteria described in Sect.~\ref{sect:PN_detection}. 

We also present the PNLF for all galaxies. In all cases, a grey, dashed histogram show the observed PNLF, and a histogram with pink solid lines indicates the PNLF corrected by completeness. Grey thick vertical bars indicate the errors. The analytic fits with a fixed and a variable bright cut-off are indicated with a dash-dotted light blue curve and a solid dark blue one, respectively. 

Finally, we present the PN number density (pink crosses), offset to match the S\'ersic fit to the surface brightness profile of each galaxy, shown with a solid black line.

\begin{figure}
    \centering
    \includegraphics[width=\linewidth]{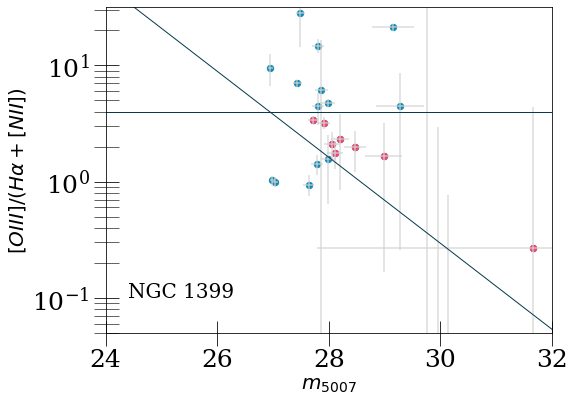}
    \caption{$\rm{[\ion{O}{iii}]}/(H{\alpha}+\rm{[\ion{N}{ii}]})$ line ratios as a function of [\ion{O}{iii}] apparent magnitude for the entire sample of objects detected in NGC\,1399 using the [\ion{O}{iii}] field. Symbols and lines as described in Figure\,\ref{fig:diagnp}}
    \label{fig:1399-diag}
\end{figure}

\begin{figure}
    \centering
    \includegraphics[width=\linewidth]{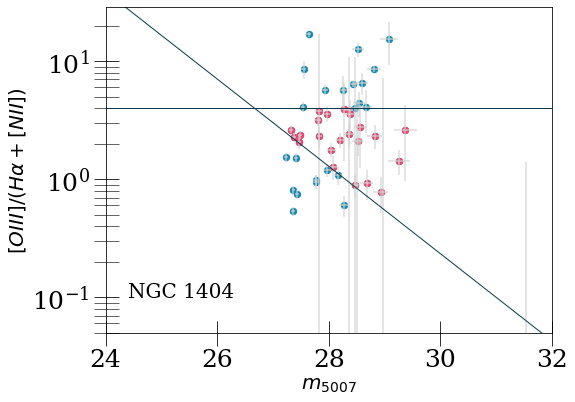}
        \caption{$\rm{[\ion{O}{iii}]}/(H{\alpha}+\rm{[\ion{N}{ii}]})$ line ratios as a function of [\ion{O}{iii}] apparent magnitude for the entire sample of objects detected in NGC\,1404 using the [\ion{O}{iii}] field. Symbols and lines as described in Figure\,\ref{fig:diagnp}}
    \label{fig:1404-diag}
\end{figure}

\begin{figure}
    \centering
    \includegraphics[width=\linewidth]{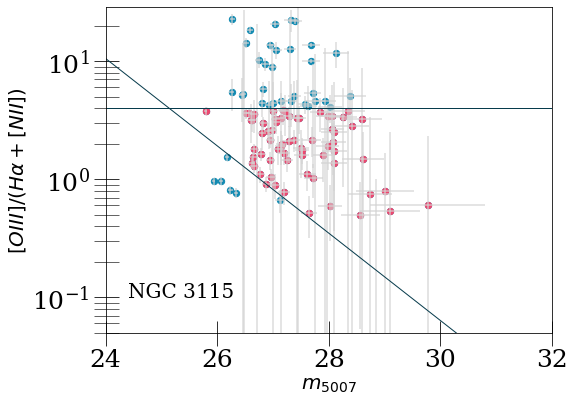}
        \caption{$\rm{[\ion{O}{iii}]}/(H{\alpha}+\rm{[\ion{N}{ii}]})$ line ratios as a function of [\ion{O}{iii}] apparent magnitude for the entire sample of objects detected in NGC\,3115 using the [\ion{O}{iii}] field. Symbols and lines as described in Figure\,\ref{fig:diagnp}}
    \label{fig:3115-diag}
\end{figure}

\begin{figure}
    \centering
    \includegraphics[width=\linewidth]{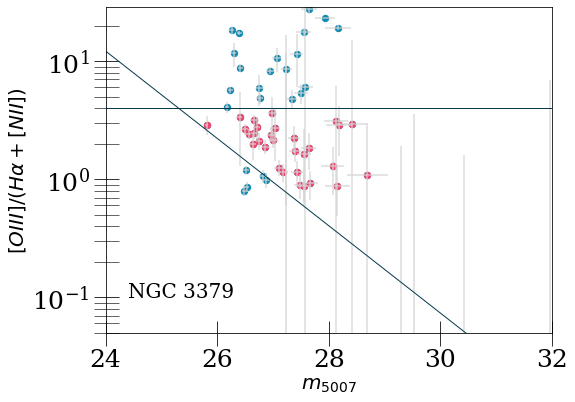}
        \caption{$\rm{[\ion{O}{iii}]}/(H{\alpha}+\rm{[\ion{N}{ii}]})$ line ratios as a function of [\ion{O}{iii}] apparent magnitude for the entire sample of objects detected in NGC\,3379 using the [\ion{O}{iii}] field. Symbols and lines as described in Figure\,\ref{fig:diagnp}}
    \label{fig:3379-diag}
\end{figure}

\begin{figure}
    \centering
    \includegraphics[width=\linewidth]{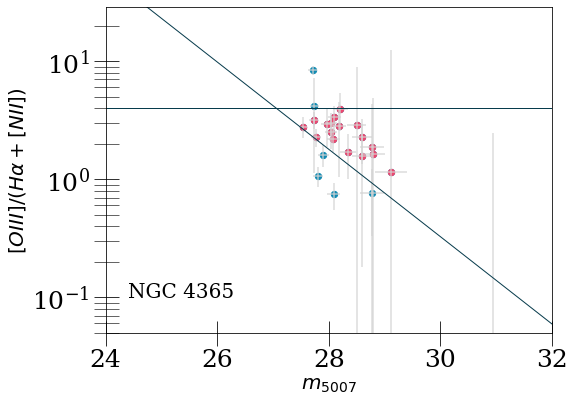}
        \caption{$\rm{[\ion{O}{iii}]}/(H{\alpha}+\rm{[\ion{N}{ii}]})$ line ratios as a function of [\ion{O}{iii}] apparent magnitude for the entire sample of objects detected in NGC\,4365 using the [\ion{O}{iii}] field. Symbols and lines as described in Figure\,\ref{fig:diagnp}}
    \label{fig:4365-diag}
\end{figure}

\begin{figure}
    \centering
    \includegraphics[width=\linewidth]{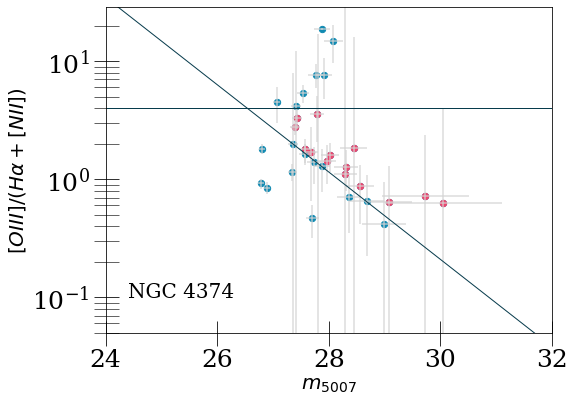}
        \caption{$\rm{[\ion{O}{iii}]}/(H{\alpha}+\rm{[\ion{N}{ii}]})$ line ratios as a function of [\ion{O}{iii}] apparent magnitude for the entire sample of objects detected in NGC\,4374 using the [\ion{O}{iii}] field. Symbols and lines as described in Figure\,\ref{fig:diagnp}}
    \label{fig:4374-diag}
\end{figure}

\begin{figure}
    \centering
    \includegraphics[width=\linewidth]{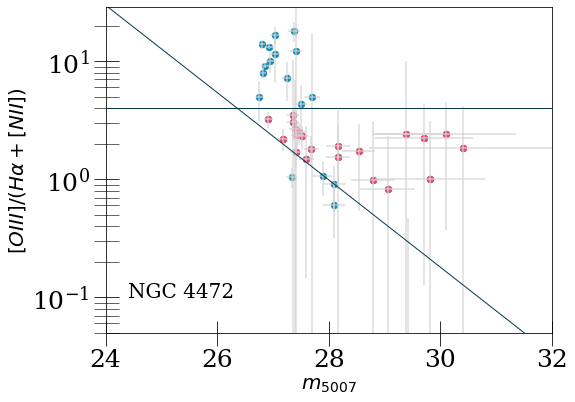}
        \caption{$\rm{[\ion{O}{iii}]}/(H{\alpha}+\rm{[\ion{N}{ii}]})$ line ratios as a function of [\ion{O}{iii}] apparent magnitude for the entire sample of objects detected in NGC\,4472 using the [\ion{O}{iii}] field. Symbols and lines as described in Figure\,\ref{fig:diagnp}}
    \label{fig:4472-diag}
\end{figure}

\begin{figure}
    \centering
    \includegraphics[width=\linewidth]{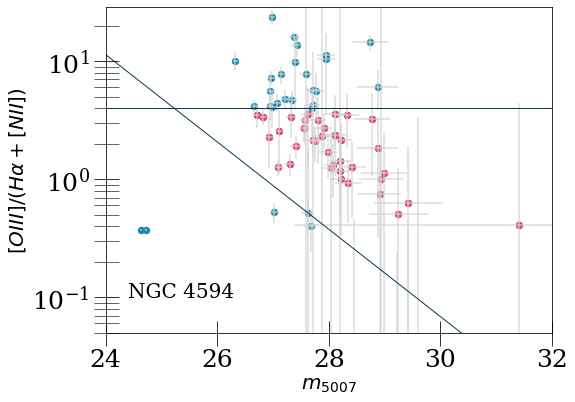}
        \caption{$\rm{[\ion{O}{iii}]}/(H{\alpha}+\rm{[\ion{N}{ii}]})$ line ratios as a function of [\ion{O}{iii}] apparent magnitude for the entire sample of objects detected in NGC\,4594 using the [\ion{O}{iii}] field. Symbols and lines as described in Figure\,\ref{fig:diagnp}}
    \label{fig:4594-diag}
\end{figure}

\begin{figure}
    \centering
    \includegraphics[width=\linewidth]{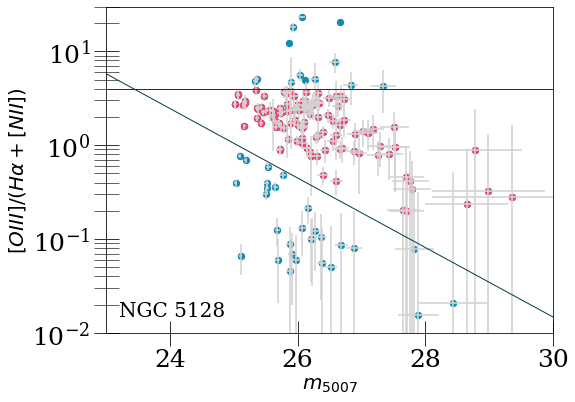}
        \caption{$\rm{[\ion{O}{iii}]}/(H{\alpha}+\rm{[\ion{N}{ii}]})$ line ratios as a function of [\ion{O}{iii}] apparent magnitude for the entire sample of objects detected in NGC\,5128 using the [\ion{O}{iii}] field. Symbols and lines as described in Figure\,\ref{fig:diagnp}}
    \label{fig:5128-diag}
\end{figure}

\begin{figure}
    \centering
    \includegraphics[width=\linewidth]{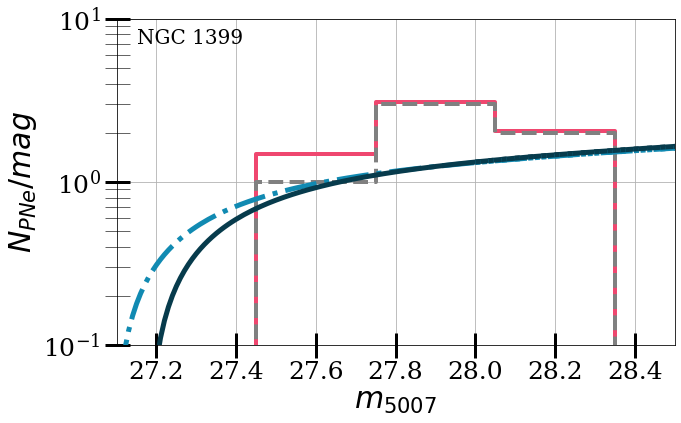}
    \caption{PNLF for PNe detected in NGC\,1399, symbols and lines as described in Figure\,\ref{fig:1387-pnlf}.}
    \label{fig:1399-pnlf}
\end{figure}

\begin{figure}
    \centering
    \includegraphics[width=\linewidth]{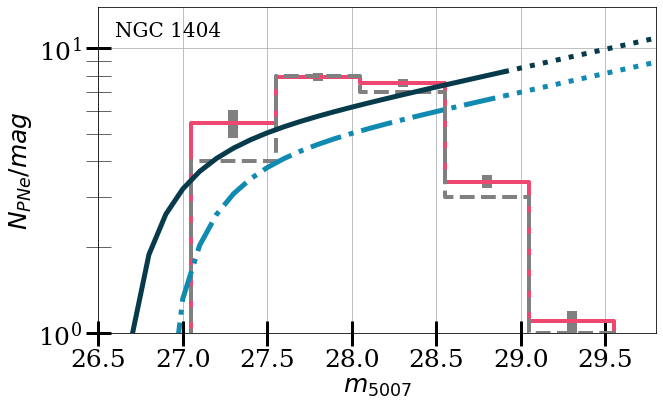}
        \caption{PNLF for PNe detected in NGC\,1404, symbols and lines as described in Figure\,\ref{fig:1387-pnlf}.}
    \label{fig:1404-pnlf}
\end{figure}

\begin{figure}
    \centering
    \includegraphics[width=\linewidth]{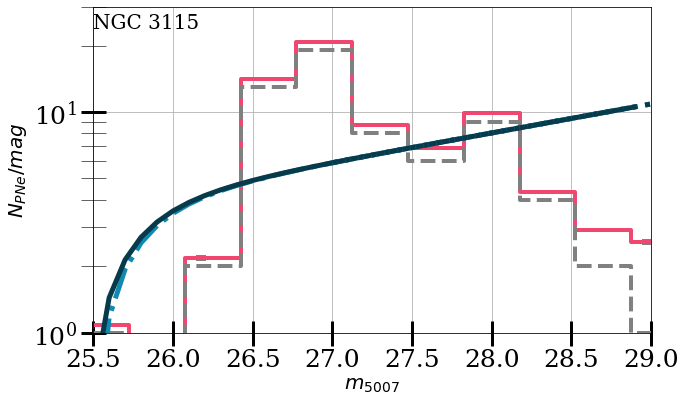}
        \caption{PNLF for PNe detected in NGC\,3115, symbols and lines as described in Figure\,\ref{fig:1387-pnlf}.}
    \label{fig:3115-pnlf}
\end{figure}

\begin{figure}
    \centering
    \includegraphics[width=\linewidth]{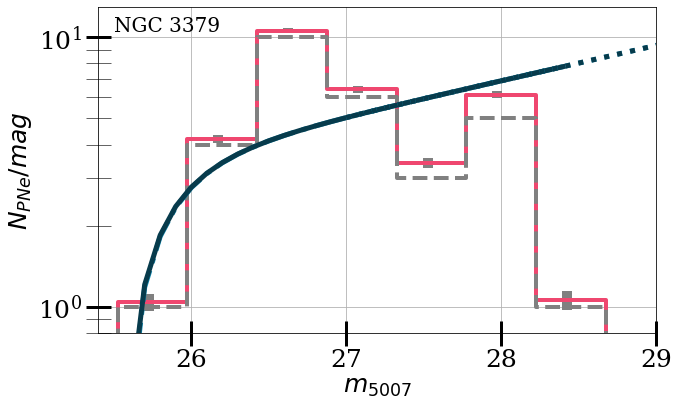}
        \caption{PNLF for PNe detected in NGC\,3379, symbols and lines as described in Figure\,\ref{fig:1387-pnlf}.}
    \label{fig:3379-pnlf}
\end{figure}

\begin{figure}
    \centering
    \includegraphics[width=\linewidth]{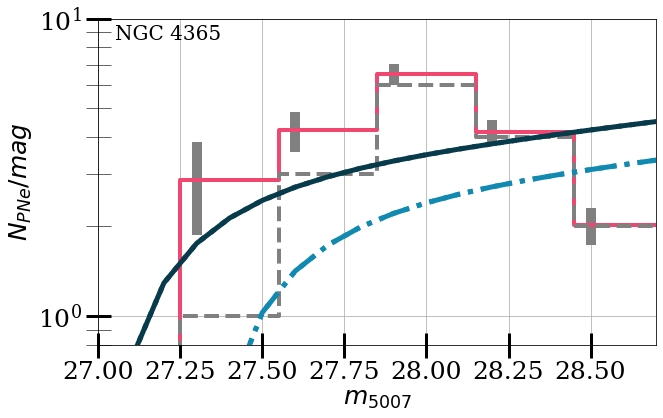}
        \caption{PNLF for PNe detected in NGC\,4365, symbols and lines as described in Figure\,\ref{fig:1387-pnlf}.}
    \label{fig:4365-pnlf}
\end{figure}

\begin{figure}
    \centering
    \includegraphics[width=\linewidth]{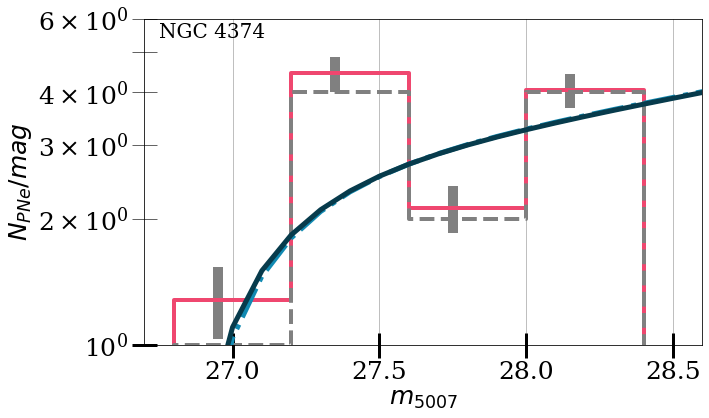}
        \caption{PNLF for PNe detected in NGC\,4374, symbols and lines as described in Figure\,\ref{fig:1387-pnlf}.}
    \label{fig:4374-pnlf}
\end{figure}

\begin{figure}
    \centering
    \includegraphics[width=\linewidth]{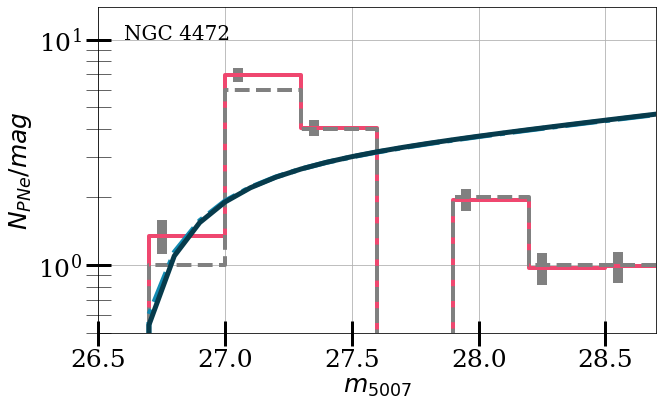}
        \caption{PNLF for PNe detected in NGC\,4472, symbols and lines as described in Figure\,\ref{fig:1387-pnlf}.}
    \label{fig:4472-pnlf}
\end{figure}

\begin{figure}
    \centering
    \includegraphics[width=\linewidth]{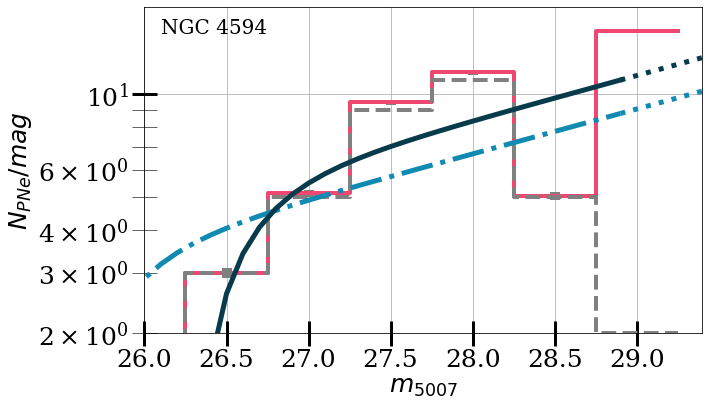}
        \caption{PNLF for PNe detected in NGC\,4594, symbols and lines as described in Figure\,\ref{fig:1387-pnlf}.}
    \label{fig:4594-pnlf}
\end{figure}

\begin{figure}
    \centering
    \includegraphics[width=\linewidth]{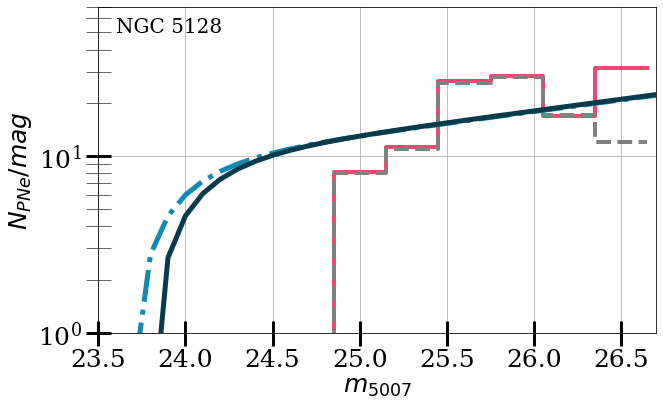}
        \caption{PNLF for PNe detected in NGC\,5128, symbols and lines as described in Figure\,\ref{fig:1387-pnlf}.}
    \label{fig:5128-pnlf}
\end{figure}

\begin{figure}
    \centering
    \includegraphics[width=\linewidth]{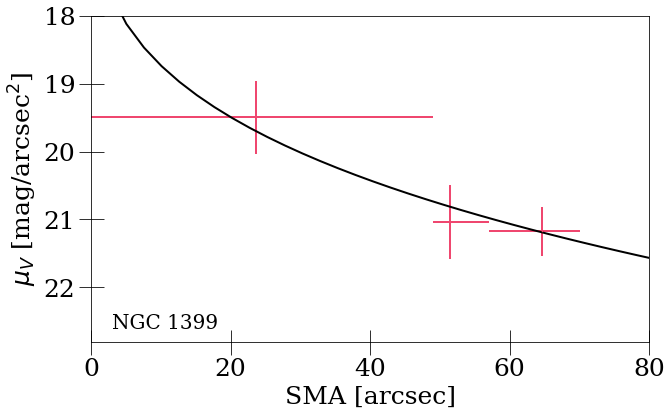}
    \caption{S\'ersic fit to the surface brightness profile for NGC\,1399 in the $V$-band shown in black solid line. Pink crosses indicate the PNe density offset by the fitted constant $\mu_{0}$, with the errorbars signaling the size of the bin in the x-axis and the density error in the y-axis}
    \label{fig:1399-surfbright}
\end{figure}

\begin{figure}
    \centering
    \includegraphics[width=\linewidth]{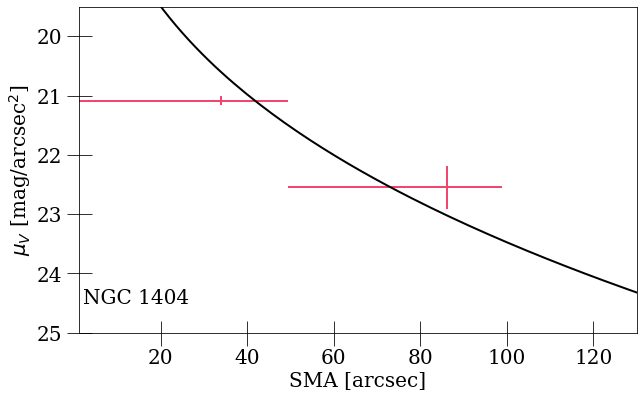}
        \caption{S\'ersic fit to the surface brightness profile for NGC\,1404 in the $V$-band shown in black solid line. Pink crosses indicate the PNe density offset by the fitted constant $\mu_{0}$, with the errorbars signaling the size of the bin in the x-axis and the density error in the y-axis}
    \label{fig:1404-surfbright}
\end{figure}

\begin{figure}
    \centering
    \includegraphics[width=\linewidth]{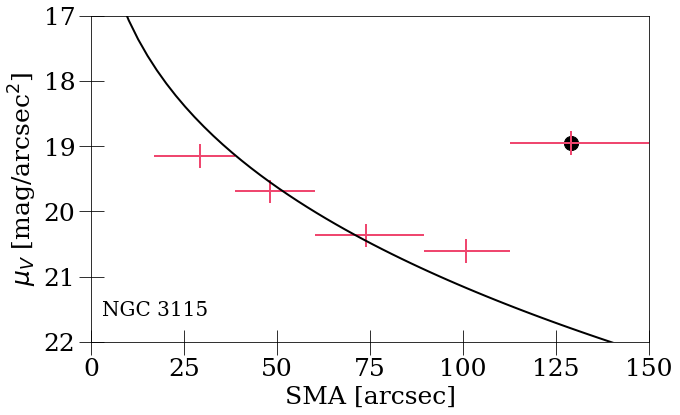}
        \caption{S\'ersic fit to the surface brightness profile for NGC\,3115 in the $V$-band shown in black solid line. Pink crosses indicate the PNe density offset by the fitted constant $\mu_{0}$, with the errorbars signaling the size of the bin in the x-axis and the density error in the y-axis. Black dots indicate points excluded from the fit.}
    \label{fig:3115-surfbright}
\end{figure}

\begin{figure}
    \centering
    \includegraphics[width=\linewidth]{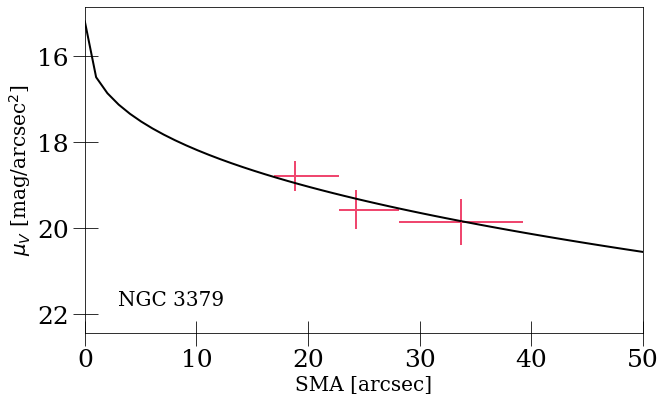}
        \caption{S\'ersic fit to the surface brightness profile for NGC\,3379 in the $V$-band shown in black solid line. Pink crosses indicate the PNe density offset by the fitted constant $\mu_{0}$, with the errorbars signaling the size of the bin in the x-axis and the density error in the y-axis}
    \label{fig:3379-surfbright}
\end{figure}

\begin{figure}
    \centering
    \includegraphics[width=\linewidth]{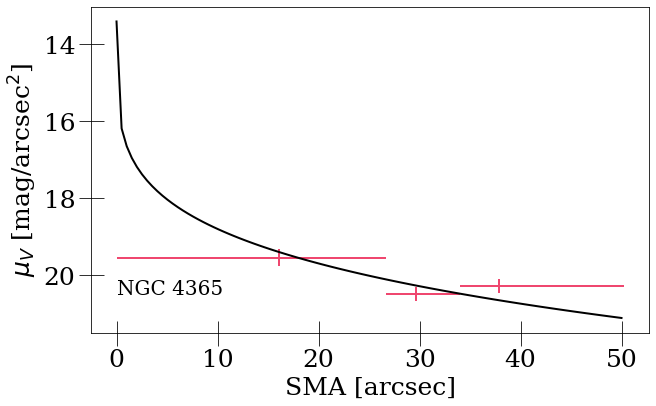}
        \caption{S\'ersic fit to the surface brightness profile for NGC\,4365 in the $V$-band shown in black solid line. Pink crosses indicate the PNe density offset by the fitted constant $\mu_{0}$, with the errorbars signaling the size of the bin in the x-axis and the density error in the y-axis}
    \label{fig:4365-surfbright}
\end{figure}

\begin{figure}
    \centering
    \includegraphics[width=\linewidth]{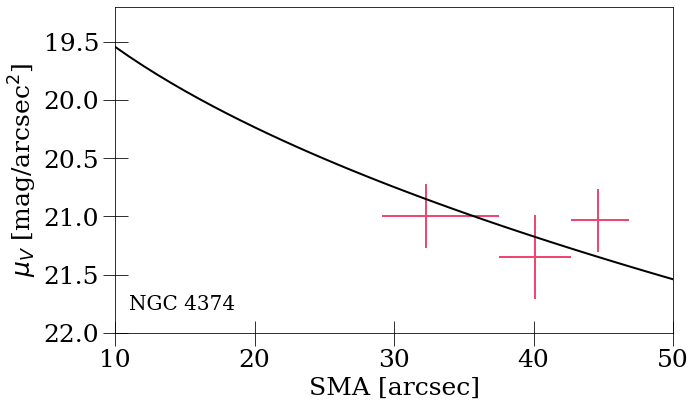}
        \caption{S\'ersic fit to the surface brightness profile for NGC\,4374 in the $V$-band shown in black solid line. Pink crosses indicate the PNe density offset by the fitted constant $\mu_{0}$, with the errorbars signaling the size of the bin in the x-axis and the density error in the y-axis}
    \label{fig:4374-surfbright}
\end{figure}

\begin{figure}
    \centering
    \includegraphics[width=\linewidth]{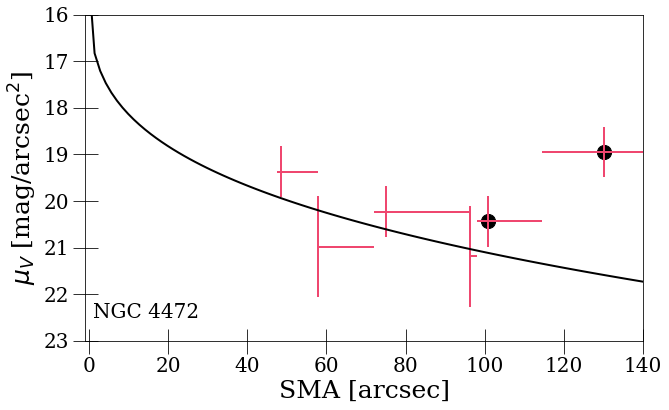}
        \caption{S\'ersic fit to the surface brightness profile for NGC\,4472 in the $V$-band shown in black solid line. Pink crosses indicate the PNe density offset by the fitted constant $\mu_{0}$, with the errorbars signaling the size of the bin in the x-axis and the density error in the y-axis. Black dots indicate points excluded from the fit.}
    \label{fig:4472-surfbright}
\end{figure}

\begin{figure}
    \centering
    \includegraphics[width=\linewidth]{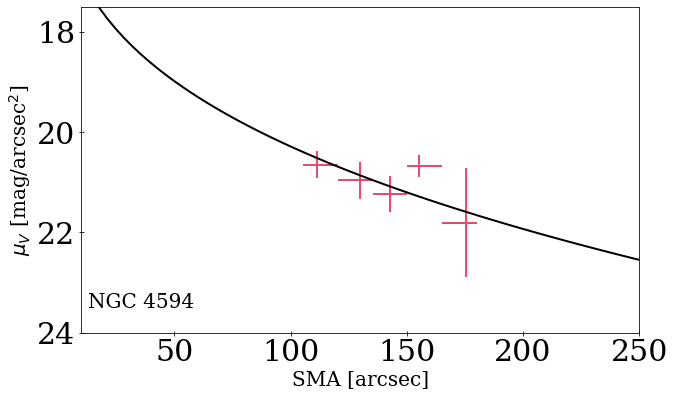}
        \caption{S\'ersic fit to the surface brightness profile for NGC\,4594 in the $V$-band shown in black solid line. Pink crosses indicate the PNe density offset by the fitted constant $\mu_{0}$, with the errorbars signaling the size of the bin in the x-axis and the density error in the y-axis}
    \label{fig:4594-surfbright}
\end{figure}

\begin{figure}
    \centering
    \includegraphics[width=\linewidth]{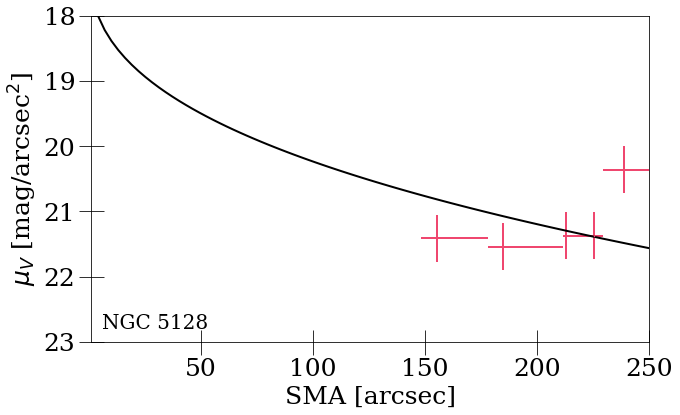}
        \caption{S\'ersic fit to the surface brightness profile for NGC\,5128 in the $V$-band shown in black solid line. Pink crosses indicate the PNe density offset by the fitted constant $\mu_{0}$, with the errorbars signaling the size of the bin in the x-axis and the density error in the y-axis}
    \label{fig:5128-surfbright}
\end{figure}

%%%%%%%%%%%%%%%%%%%%%%%%%%%%%%%%%%%%%%%%%%%%%%%%%%

% Don't change these lines
\bsp	% typesetting comment
\label{lastpage}
\end{document}